\documentclass{emulateapj}
\usepackage{color}

\slugcomment{Received 2015.05.30; Accepted}

\shorttitle{X-ray property of short gamma-ray bursts}
\shortauthors{Kagawa et al.}

\begin{document}


\title{
X-raying Extended emission and rapid decay of short gamma-ray bursts}


\author{
Yasuaki Kagawa\altaffilmark{1},
Daisuke Yonetoku\altaffilmark{1},
Tatsuya Sawano\altaffilmark{1},
Asuka Toyanago\altaffilmark{1},
Takashi Nakamura\altaffilmark{2},
Keitaro Takahashi\altaffilmark{3}, 
Kazumi Kashiyama\altaffilmark{4}, 
and 
Kunihito Ioka\altaffilmark{5,6}
}
\email{kagawa@astro.s.kanazawa-u.ac.jp}
\email{yonetoku@astro.s.kanazawa-u.ac.jp}
\altaffiltext{1}{College of Science and Engineering, 
School of Mathematics and Physics,
Kanazawa University, Kakuma, Kanazawa, Ishikawa 920-1192, Japan}
\altaffiltext{2}{Department of Physics, Kyoto University, Kyoto
606-8502, Japan}
\altaffiltext{3}{Faculty of Science, Kumamoto University, Kurokami,
Kumamoto, 860-8555, Japan}
\altaffiltext{4}{Einstein fellow --- Department of Astronomy; Department of Physics; 
Theoretical Astrophysics Center; University of California, 
Berkeley,Berkeley, CA 94720, USA}
\altaffiltext{5}{Theory Center, Institute of Particle and Nuclear Studies, KEK,
Tsukuba 305-0801, Japan}
\altaffiltext{6}{Department of Particle and Nuclear Physics,
SOKENDAI (the Graduate University for Advanced Studies),
Tsukuba 305-0801, Japan}

\begin{abstract}
Extended emission is a mystery in short gamma-ray bursts (SGRBs).
By making time resolved spectral analyses of brightest nine events observed by {\it Swift} XRT,
we obviously classify the early X-ray emission of SGRBs into two types.
One is the extended emission with exponentially rapid decay,
which shows significant spectral softening 
during hundreds seconds since the SGRB trigger
and is also detected by {\it Swift}-BAT.
The other is a dim afterglow only showing power-law decay over $10^4$ s.
The correlations between the temporal decay and spectral indices 
of the extended emissions
are inconsistent with the $\alpha$-$\beta$ correlation expected for
the high-latitude curvature emission from a uniform jet.
The observed too-rapid decay suggests the emission from a photosphere or a patchy surface,
and manifests the stopping central engine 
via such as magnetic reconnection at the black hole.
\end{abstract}


\section{Introduction}
Short Gamma-Ray Burst (SGRB) is a sub-category of gamma-ray burst 
phenomena. The SGRB lightcurve is composed of an intense prompt 
emission with short time duration less than 2~sec and an extended 
soft X-ray emission lasting about 100~sec in some cases. 
The integrated energy of both emissions are almost comparable, 
which motivates us to make detailed studies of the X-ray properties.

The origin of SGRB is still in debate. 
A major candidate is a coalescence of compact objects, 
such as neutron stars and black holes~
\citep{paczynski1986, eichler1989}. 
Recently, in the afterglow of GRB~130603B, re-brightening in 
red or near infrared band, a so-called macronova (or kilonova),
was observed, and its physical interpretation is discussed as 
nuclear decay of neutron-rich r-process (rapid-process)
elements synthesized in the ejecta of a neutron star binary 
coalescence~\citep{tanvir2013, berger2013,berger2014}.
There are also other possibilities that the macronova is 
energized by the extended activity of the central engine 
\citep{kisaka+15} and/or dust emission \citep{takami+14}.

In the compact merger scenario, the SGRB prompt emission is 
powered by a relativistic jet launched from the remnant compact 
object surrounded by a massive disk. The central engine would be 
black holes 
\citep{fan2005,rosswog2007,lee2009,barkov2011,nakamura2014,kisaka2015}, 
or rapidly-spinning strongly-magnetized neutron stars 
\citep[millisecond magnetars;][]{usov1992,zhang2001,gao2006,metzger2008,
bucciantini2012,gompertz2013,zhang2013}, 
depending on the type of coalescing binaries and the equation of 
state of neutron star matter \citep{hotokezaka2011,kyutoku+13,kyutoku+15}.
The origin of the extended emission is rather puzzling; 
the observed duration is longer than the typical accretion time of 
the disk. It has been proposed that the longer activity can be 
powered by fallback accretion of tidally stripped matter
\citep{lee2009,kisaka2015} or spindown of 
magnetars~\citep{metzger2008}. 

If the compact merger scenario is the case, we expect to detect 
strong gravitational waves by the second generation 
gravitational wave observatories, 
Advanced-LIGO\footnote{http://www.ligo.caltech.edu/},
Advanced-VIRGO\footnote{http://www.ego-gw.it/index.aspx/}
and KAGRA\footnote{http://gwcenter.icrr.u-tokyo.ac.jp/en/}.
To pioneer the gravitational wave astronomy, 
synchronized observations of electromagnetic counterparts 
in multi-wavelengths are required. Wide field monitoring in 
X-ray band is an important method to discover the prompt emission 
and extended soft X-ray emission of SGRBs accompanying 
the gravitational wave detection.

This paper is constructed as follows. 
In the next section, we systematically investigate X-ray 
properties of bright 9 SGRBs. Especially, we perform 
the time resolved spectral analyses for the extended soft X-ray 
emission and X-ray afterglows. 
We show that strong spectral evolution is a common property of 
the rapid decay phase that follows the extended emission of SGRBs.
Finally, in \S~\ref{sec:discussion}, we discuss the association 
between the extended soft X-ray emission and the rapid decay, 
and their spectral and temporal properties.
We find that the decay is more rapid than the simple high-latitude emission
and discuss interesting implications.
Moreover we discuss a poplation statistics having bright extended emissions.

\section{Observations \& Data Analyses}
\label{sec:analyses}
\subsection{Data Selection}
\label{subsec:selection}
We first select 79 SGRBs with the time duration of 
$T_{90} < 2.0~{\rm sec}$ from {\it Swift} gamma-ray burst catalog 
until the end of 2014. Here $T_{90}$ is measured as the duration 
of the time interval during which 90~\% of the total observed 
counts have been detected. However, considering that several SGRBs 
show the extended soft X-ray emission lasting $\sim 100~{\rm sec}$ 
just after the short prompt emission, the above criterion may 
be insufficient to select samples of interest. 
We check the individual lightcurves and pick up events with 
an initial spike of prompt emission followed by rather gradual 
time sequence. Then, we additionally include 
possible 13~SGRB candidates reported in GCN circulars\footnote{
GRB~050724, 050911, 051227, 060717, 061006, 070714B, 
080123, 080503, 090309, 100213A, 100816A, 130716A, and 130822A}.

We use data of X-ray afterglows and/or extended X-ray emission 
observed by Burst Alert Telescope (BAT) and X-ray telescope (XRT) 
aboard {\it Swift} to investigate X-ray properties of SGRBs. 
For our purpose, we need enough number of X-ray photons to perform 
time-resolved spectral analyses. Thus, we adopt a selection criterion 
of $F \ge 10^{-11}~{\rm erg~cm^{-2}s^{-1}}$ in $2-10$~keV band 
at the XRT observation start time. Hereafter, we treat selected 
9 samples, GRB~050724, 051221A, 060313, 070714B, 080503, 090510, 
100702A, 120804A, and 130603B.

\subsection{{\it Swift}-BAT lightcurves}
\label{subsec:bat-lightcurve}
\begin{figure}
\begin{center}
\includegraphics[angle=0,scale=0.20]{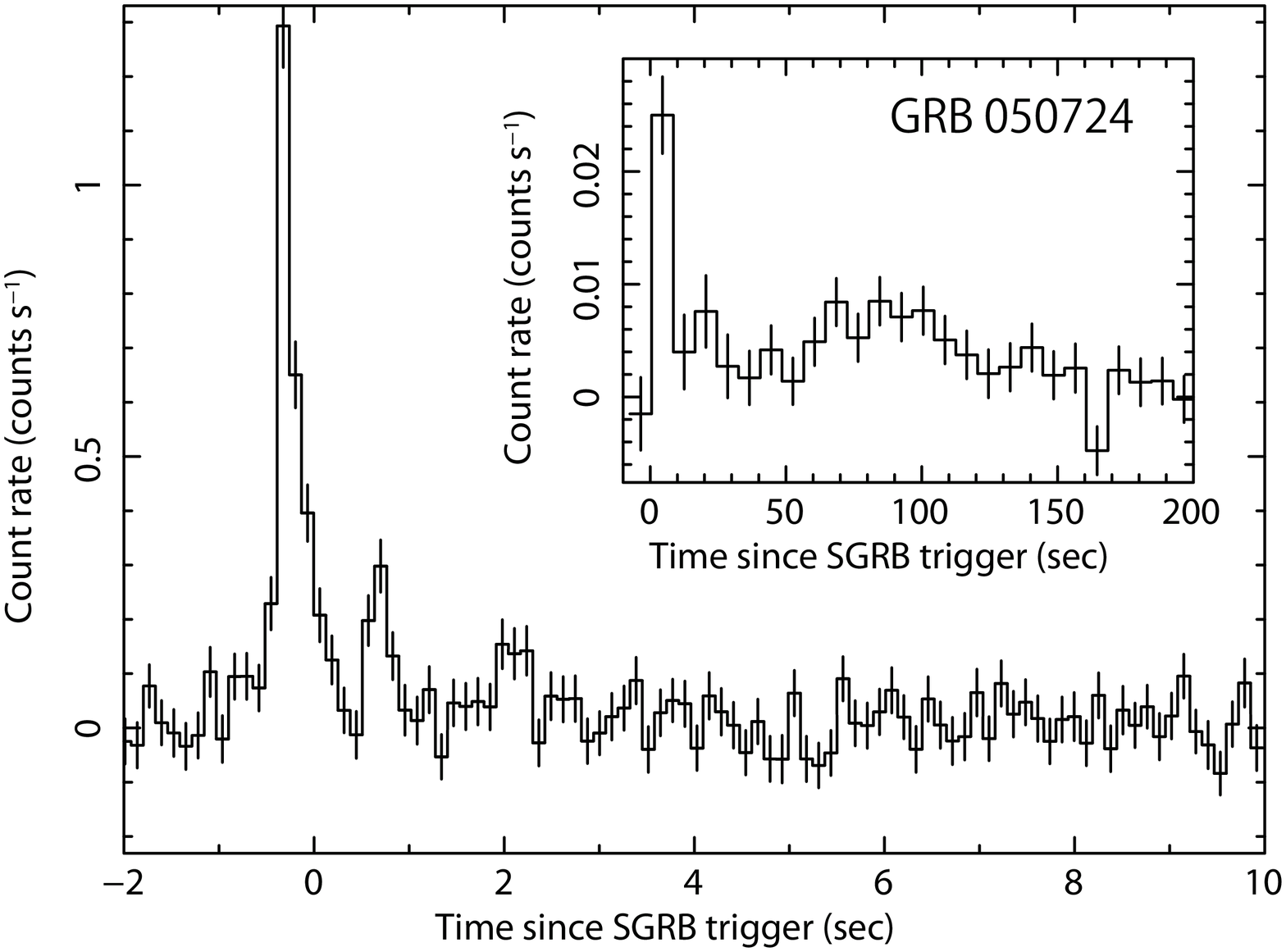}
\includegraphics[angle=0,scale=0.20]{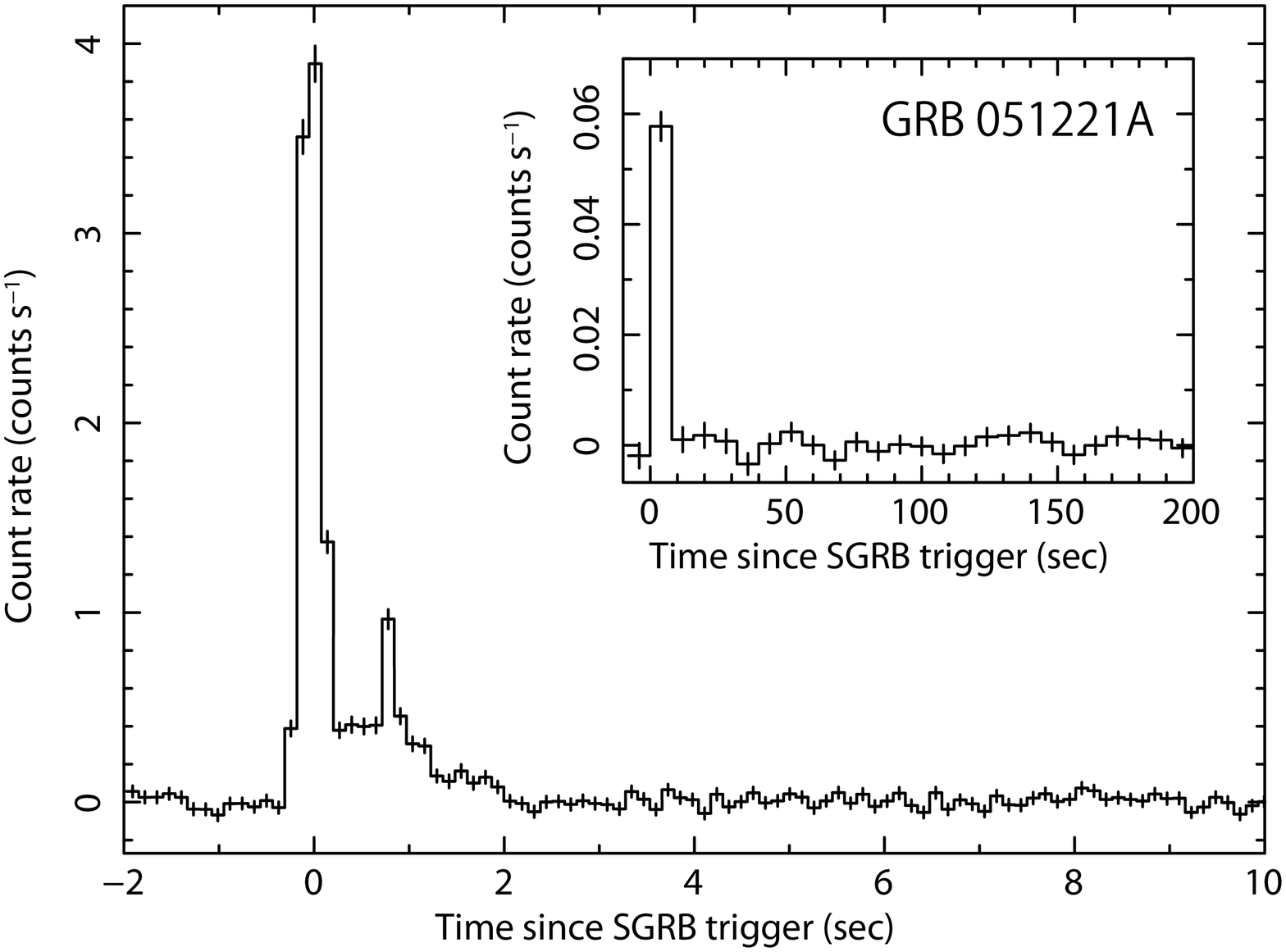}
\includegraphics[angle=0,scale=0.20]{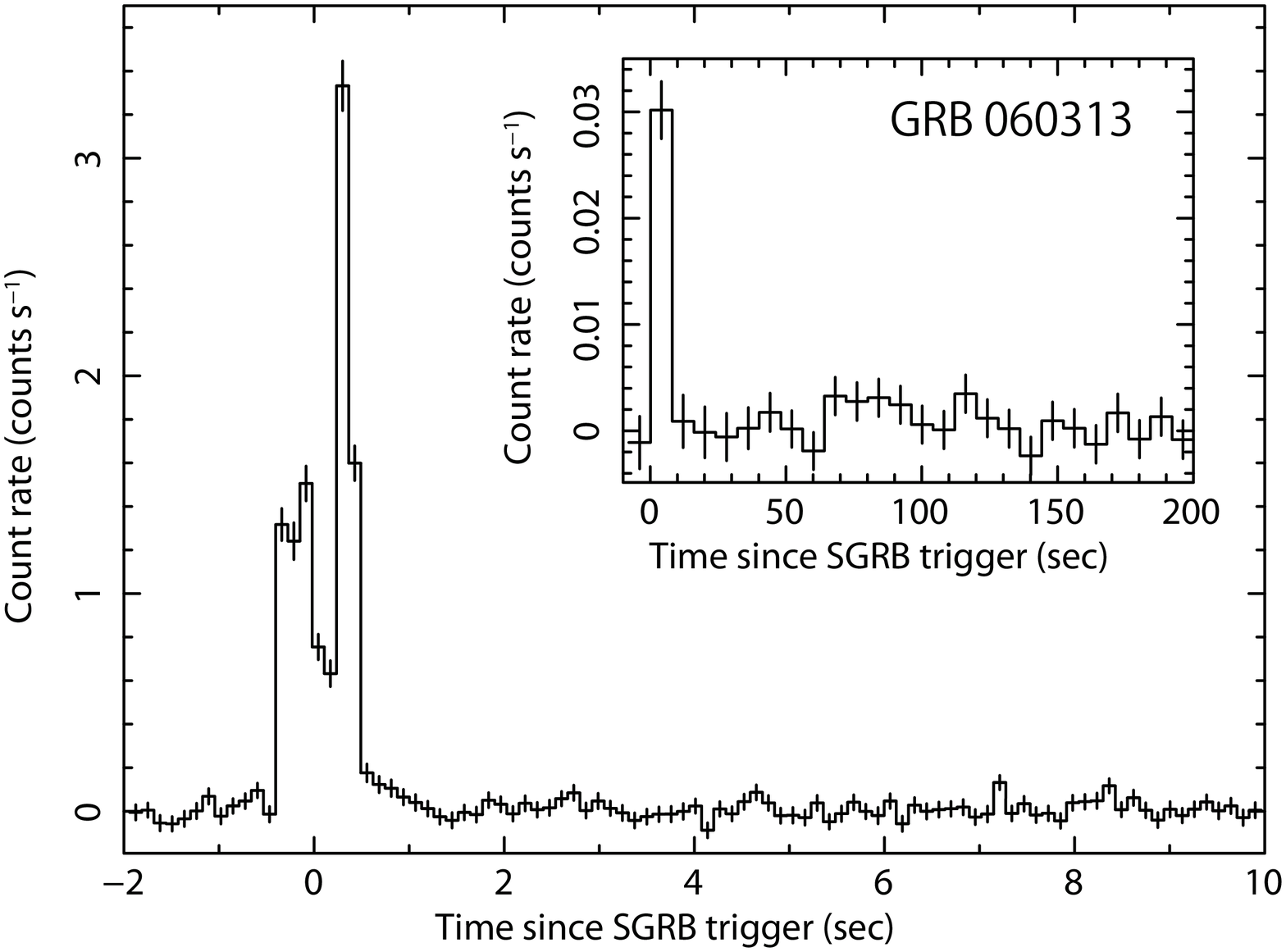}
\includegraphics[angle=0,scale=0.20]{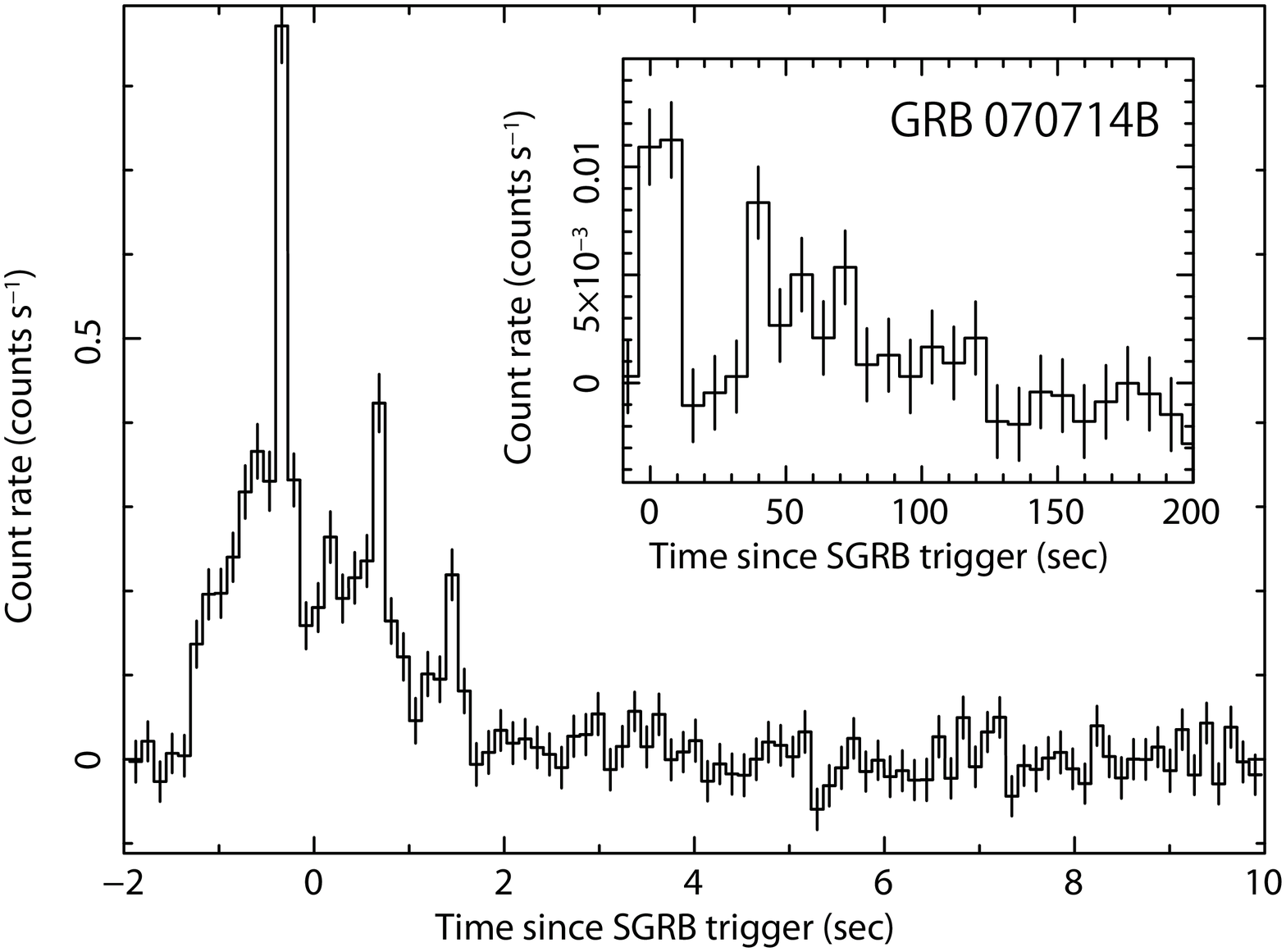}
\includegraphics[angle=0,scale=0.20]{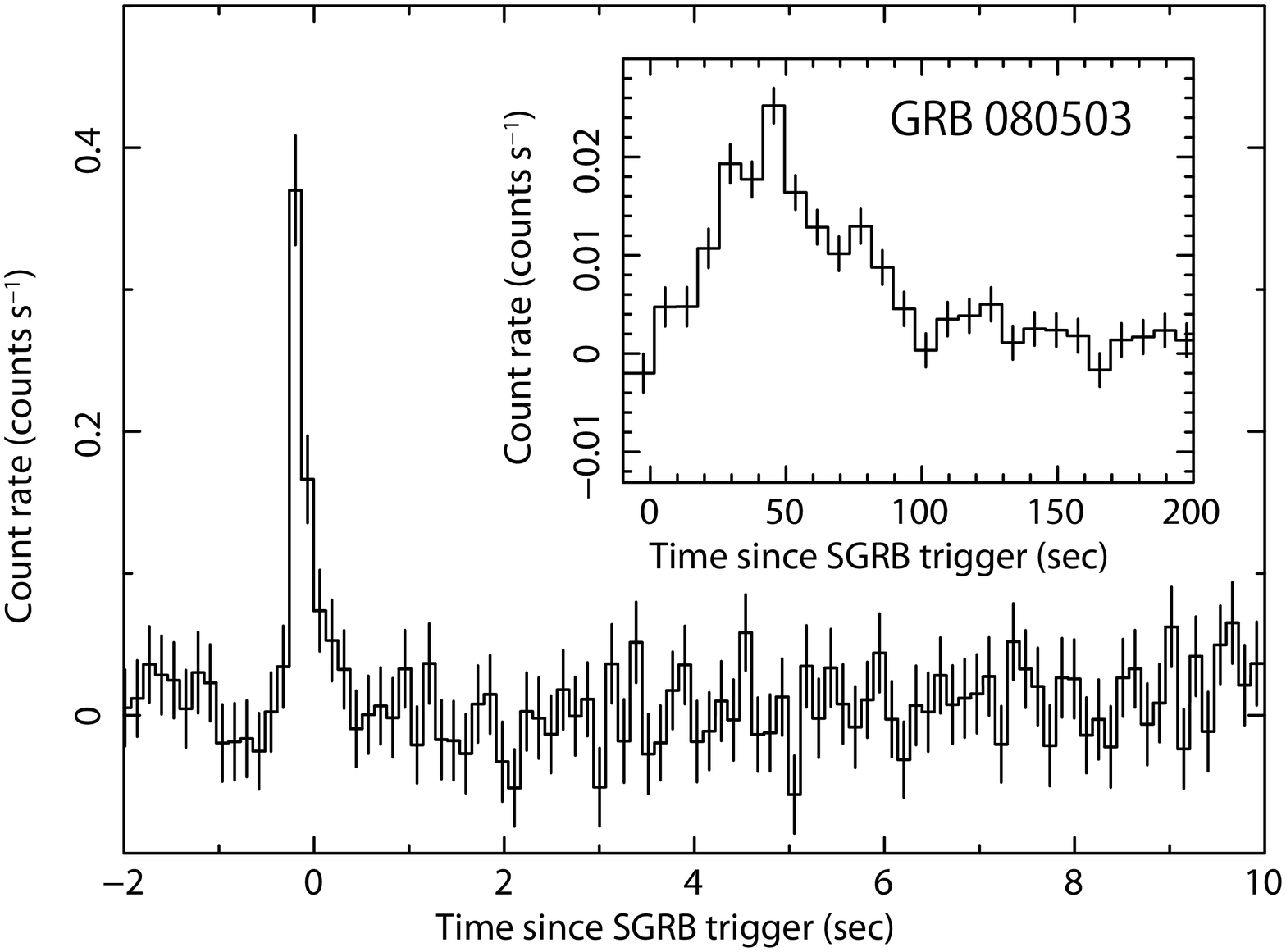}
\includegraphics[angle=0,scale=0.20]{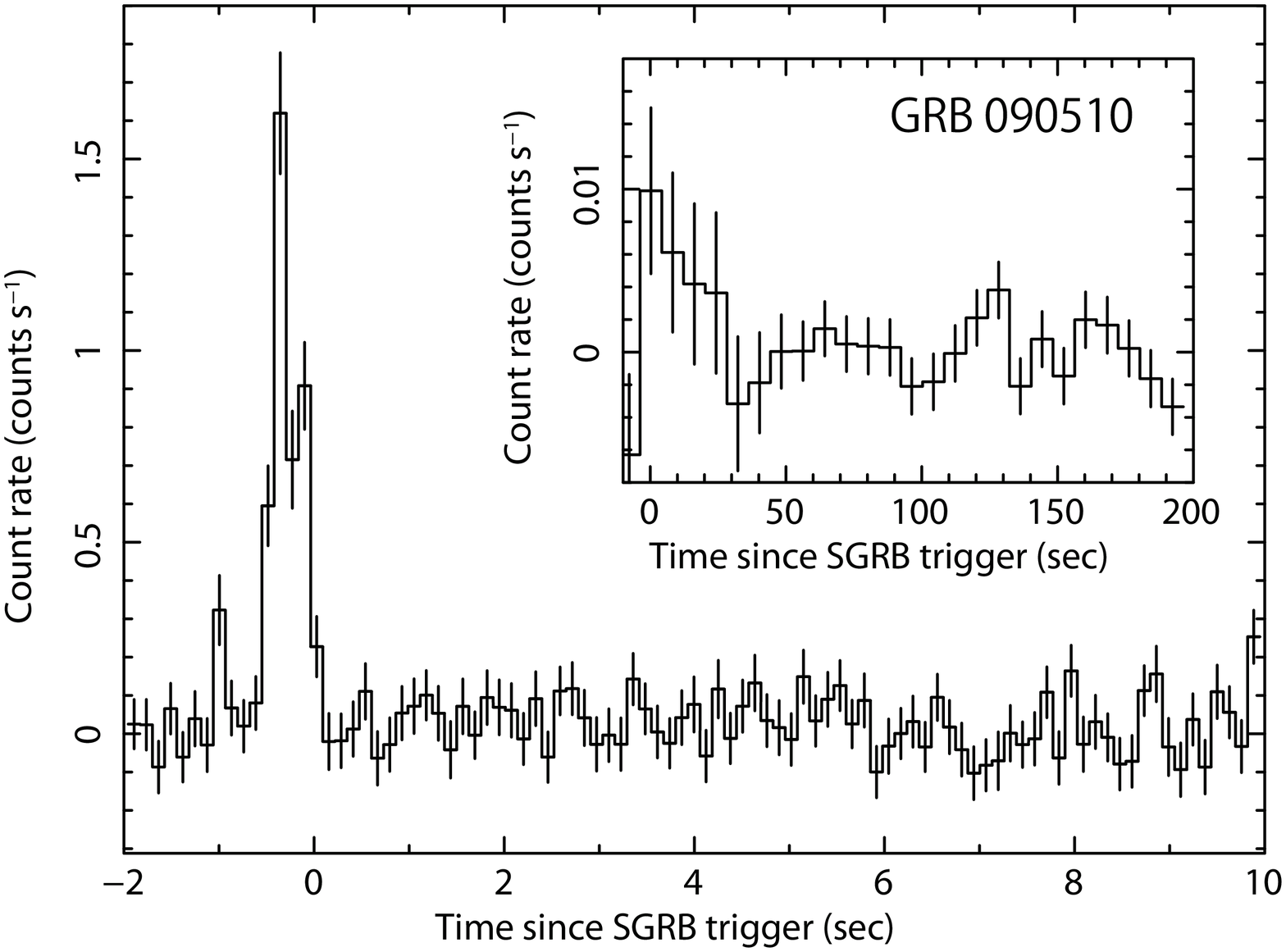}
\includegraphics[angle=0,scale=0.20]{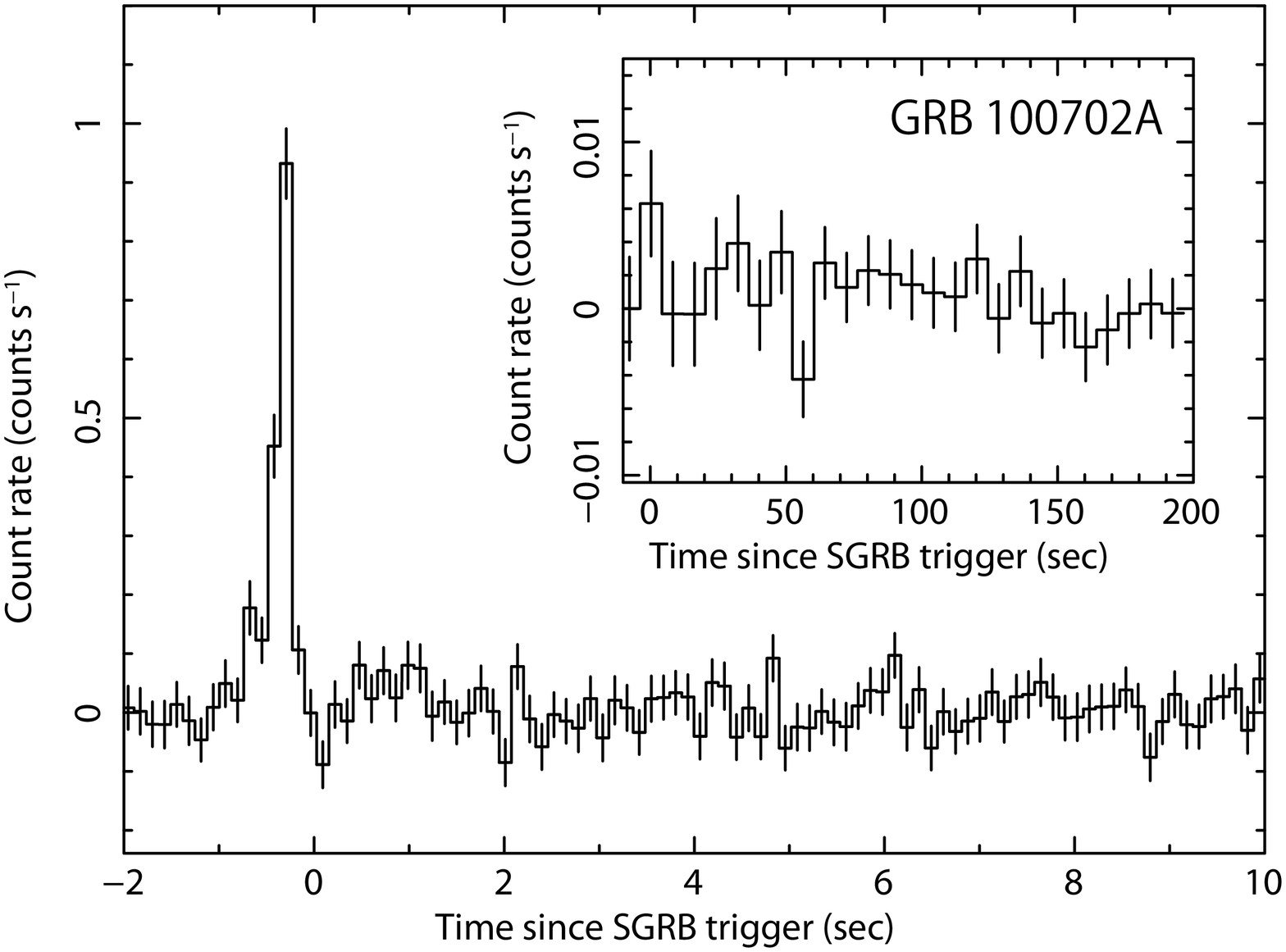}
\includegraphics[angle=0,scale=0.20]{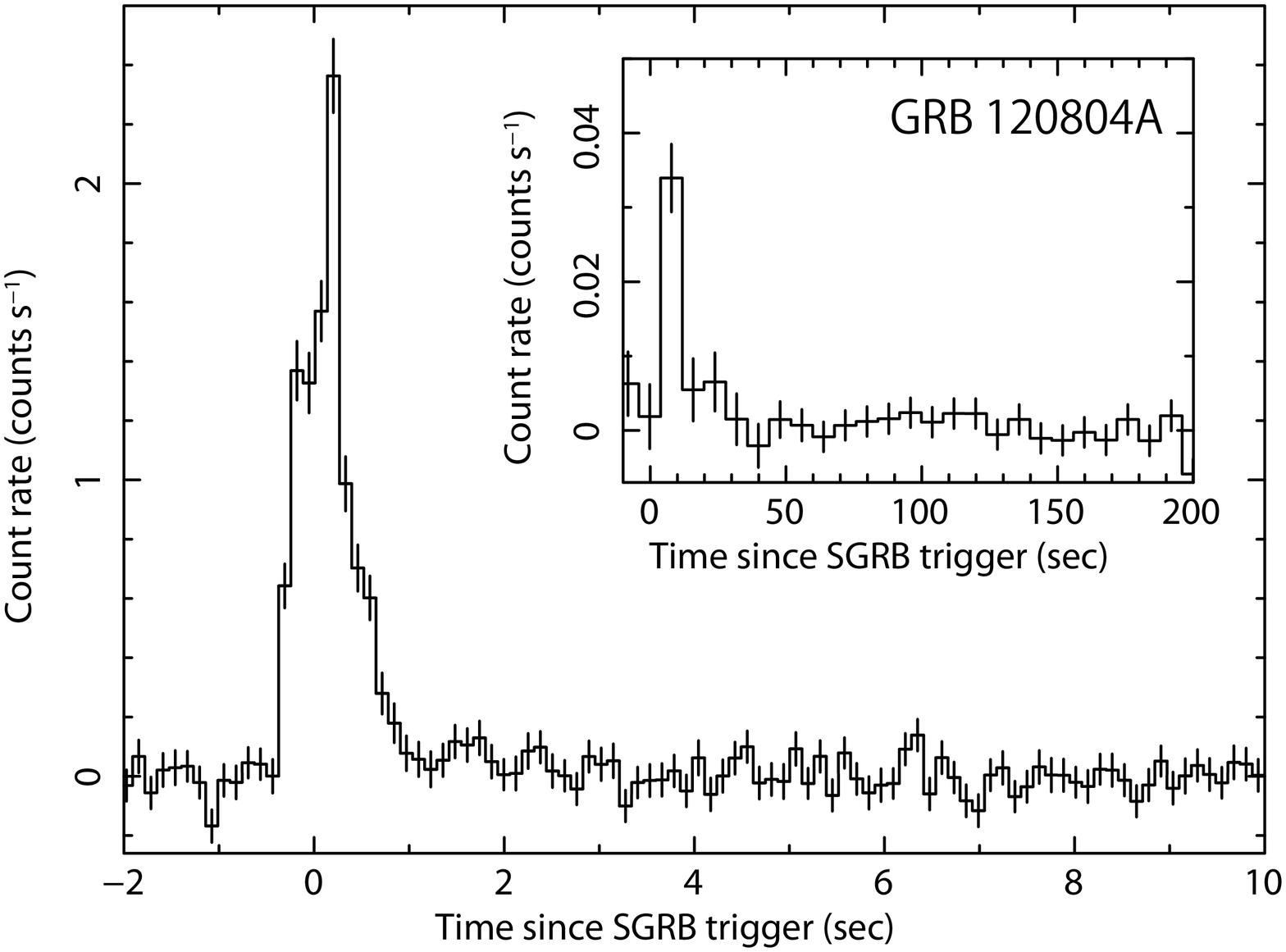}
\includegraphics[angle=0,scale=0.20]{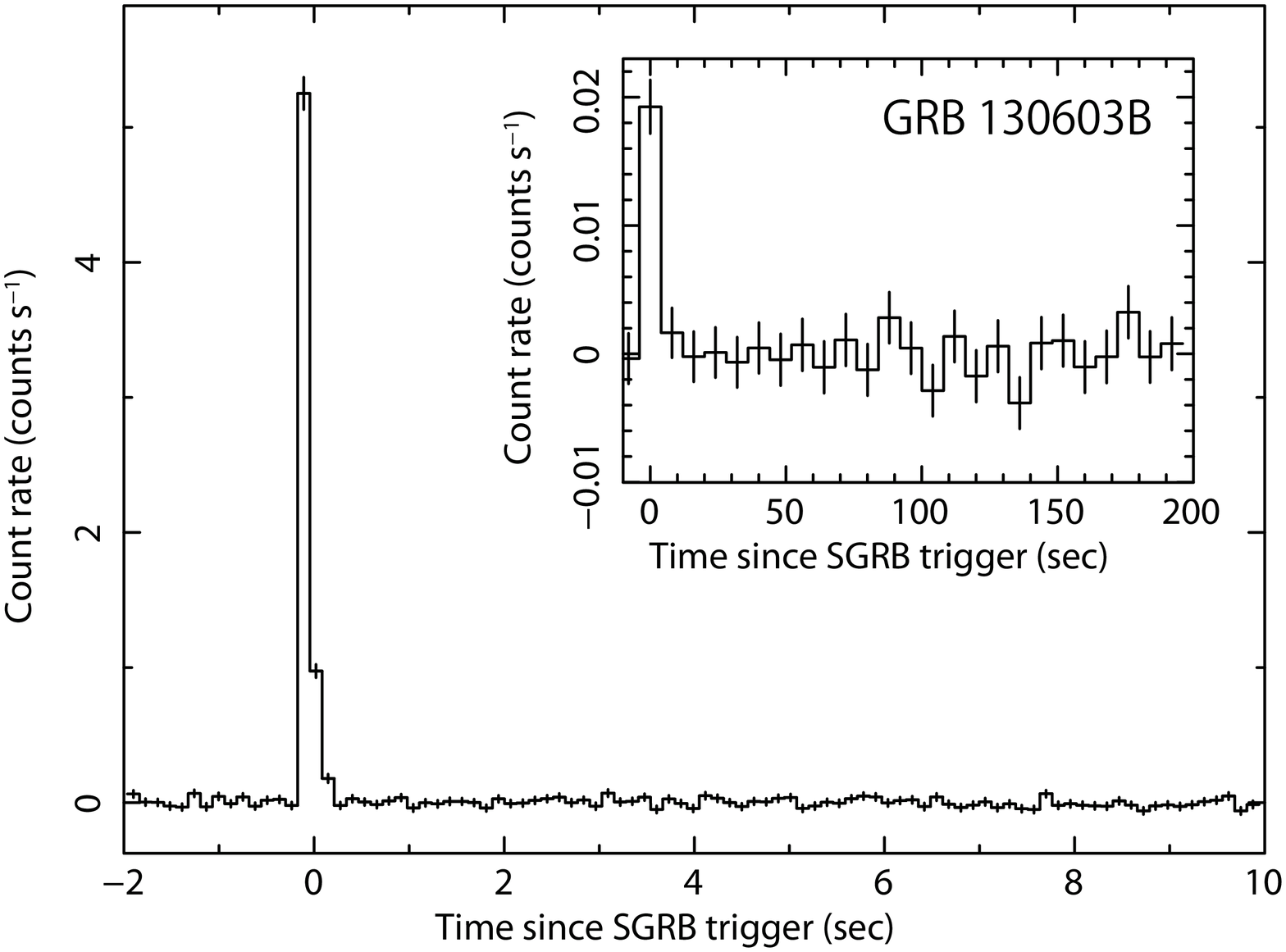}
\caption{{\it Swift}-BAT lighrcurves of bright 9 SGRBs.
The main panels show lightcurves of prompt emission with 
128~msec time resolution in 15--150~keV band. 
The inserted panels also show 
the following long-lasting
X-ray emission up to 200~sec since SGRB trigger with 
8~sec time resolution in 15--25~keV band. We 
can clearly confirm
extended X-ray emission for 3 events 
(GRB~050724, 070714B and 080503).
\label{bat-lc}}
\end{center}
\end{figure}

In Figure~\ref{bat-lc}, we show the SGRB lightcurves observed 
with BAT. In Figure~\ref{bat-lc}, each panel shows two kinds of information,
i.e., main panels show the initial spike of prompt emission 
with 128~msec time resolution in $15-150$~keV band, 
and inserted panels show the following hard X-ray emission up to
200~sec since the SGRB trigger time with 8~sec time resolution in 
$15-25$~keV band.

We adopted a null-flux constant model for data with 8~sec time bin 
in a time between $8-200$~sec since SGRB trigger time 
(except for the first 8~sec to avoid the contribution from 
prompt emission). We summarized the fitting results 
in Table~\ref{tbl:table1}. As also clearly seen in 
Figure~\ref{bat-lc}, the extended X-ray emission is detected 
in 3 samples of GRB~050724, 070714B, and 080503, with 
$T_{90}$ duration of 98.7~sec, 65.2~sec and 274.9~sec, 
respectively. 

According to these analyses, the other 6 events 
are fully consistent with no bright extended X-ray emission 
after the initial spike in BAT data.
We searched the extended X-ray emission with several 
time durations from 8~sec since the trigger time.
However we failed to detect any obvious indication 
of extended X-ray emission in these 6 events.

\subsection{Spectral Analyses of XRT data}
\label{subsec:xrt-spectra}
At first, we extracted an X-ray signal within the image region of 
$40 \times 30$ rectangular pixels with rotation angle of spacecraft 
for windowed timing (WT) mode data, 
and 20 pixels in radius (corresponding to $\sim$ 47 arcsec) 
for photon counting (PC) mode data. 
These are recommended region size described in the {\it Swift} 
XRT software guide. We extracted a background signal 
from the image region with no X-ray sources (under the sensitivity 
of XRT). The region size is the rectangular with $30 \times 30$ 
pixels for WT mode, and the circle as large as possible 
(at least 20 pixels) for PC mode data. The source and background 
regions do not overlap each other.

After that, we performed time resolved spectral analyses for 
both WT and PC data of selected 9~SGRBs. 
To conserve the uniform statistical uncertainty for each spectrum, 
we divided the entire data into several time bins to keep 
the same number of photons (about 512 photons for 
WT mode, and 256 photons for PC mode). 
Then we obtained at least 4 spectra for each SGRBs.

We adopted two kinds of spectral models, i.e. a power-law and 
a blackbody function.
We include galactic and extra-galactic 
column densities (``phabs'' model) 
for each time resolved data. The exact formulas are as follows; 
\begin{eqnarray}
\label{eq:power-law}
N(E) =  
\exp \Bigr(- (N_{H}^{gal} + N_{H}^{ext}) \sigma(E) \Bigl) \times
K \Bigr( \frac{E}{1~{\rm keV}} \Bigl)^{- \Gamma},
\end{eqnarray}
for the power-law model, and 
\begin{eqnarray}
\label{eq:blackbody}
N(E) =  
\exp (- (N_{H}^{gal} + N_{H}^{ext}) \sigma(E)) \times
\Bigr( \frac{K \times 8.0525E^2dE}{(kT)^4[\exp(E/kT)-1]} \Bigl),
\end{eqnarray}
for the blackbody model, respectively.
Here, 
$N(E)$ is in units of $\rm{photons~cm^{-2}~s^{-1}~keV^{-1}}$.
$N_{H}^{gal}$ and $N_{H}^{ext}$ are a galactic and an extra-galactic
hydrogen column density, in units of $10^{22}~{\rm atoms~cm^{-2}}$, 
respectively, and $\sigma(E)$ is a photo-electric cross-section 
(not including Thomson scattering).
$\Gamma$ is a photon index, and $kT$ is a temperature in units of keV.
The parameters K in both functions are normalization.

\begin{figure}
\includegraphics[angle=0,scale=0.32]{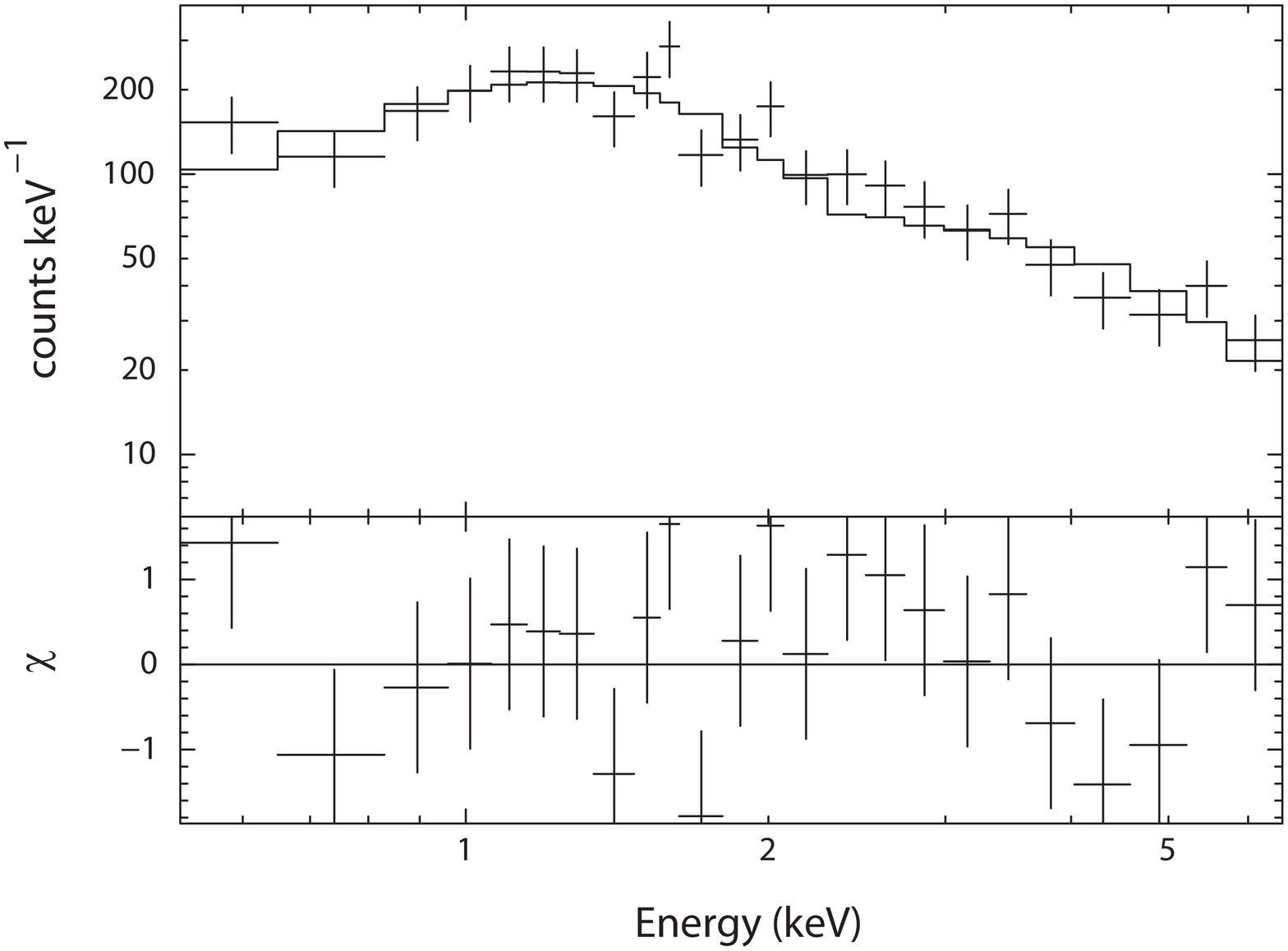}
\includegraphics[angle=0,scale=0.32]{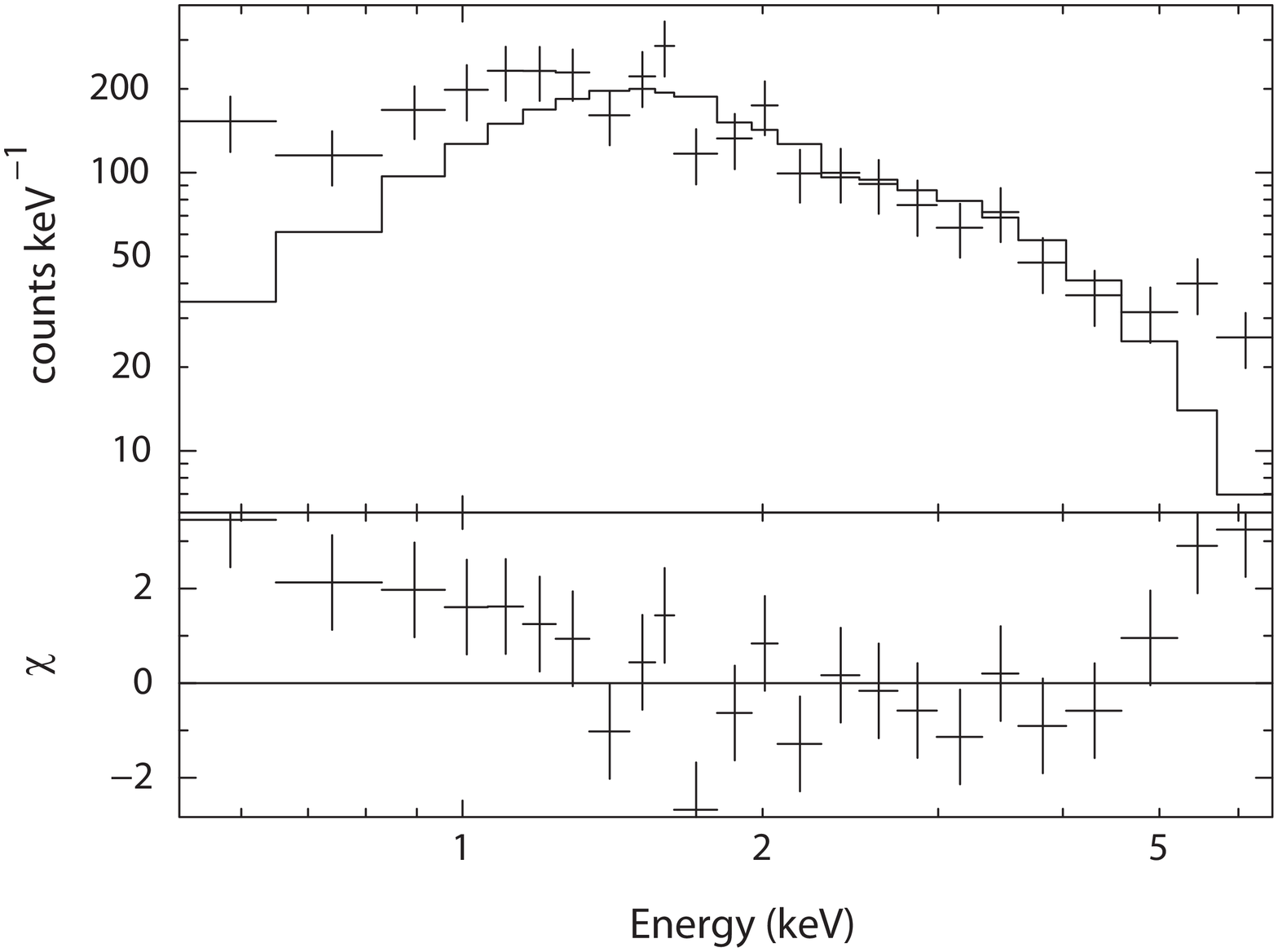}
\caption{Examples of spectral fittings of XRT data. 
Top and bottom panels are the observed spectra
(including detector response) and the fitting residuals 
of the best fit functions.
(Left) Fitting result of GRB~080503 with the absorbed 
power-law model (Equation~\ref{eq:power-law}). 
(Right) Same as the left but with the blackbody 
function (Equation~\ref{eq:blackbody}).
\label{xrt-spec}}
\end{figure}

We show representative spectra of GRB~080503 in Figure~\ref{xrt-spec}.
The left and right panels of Figure~\ref{xrt-spec} show 
the best fit results with the power-law and blackbody model, respectively.
The power-law model can well describe the entire shape of 
the observed spectrum, while the blackbody model has large 
discrepancies
between data and the model around the low- and high-energy region.

\begin{figure}
\begin{center}
\includegraphics[angle=270,scale=0.5]{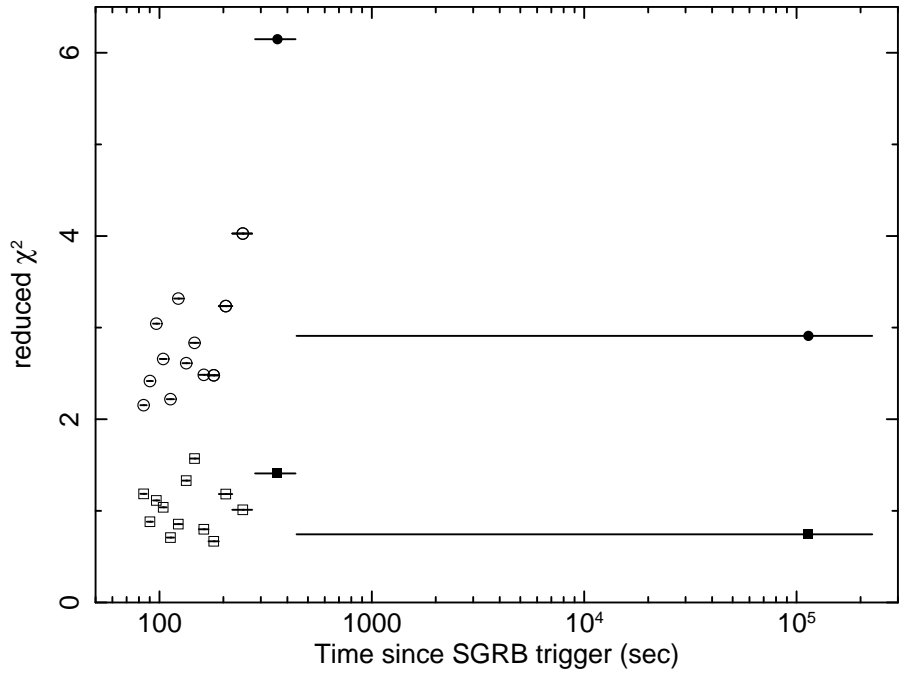}
\caption{Reduced $\chi^{2}$ distributions of the best fit 
spectral model of power-law (square) and blackbody (circle) for time resolved 
spectra of GRB~080503. The open and filled squares are 
the result of the power-law model for WT and PC mode data, 
respectively. The open and filled circles are the same but 
for the blackbody model, respectively. 
\label{pl-bb}}
\end{center}
\end{figure}

In Figure~\ref{pl-bb}, we show reduced $\chi^{2}$ distribution as 
a function of time since GRB~080503 trigger. 
The reduced $\chi^{2}$ values of 
the power-law model (open and filled squares for WT and PC mode) 
and blackbody model (open and filled circles for WT and PC mode) 
systematically locate around 1 and 2 throughout the entire epoch, 
respectively. Therefore we can conclude that the observed spectra 
can be described by the power-law model, and the blackbody model 
is not suitable for the X-ray spectra after $\sim 100$ seconds 
of SGRBs. In the other 8 events, all spectra can be described by 
the power-law model well.

\subsection{Spectral Softening}
\label{subsec:softening}
We summarized the fitting results of spectral parameters,
the energy flux in units of $10^{-12}~{\rm erg~cm^{-2}s^{-1}}$ 
in $2-10$~keV band (top panel), photon index $\Gamma$ (middle panel), 
and the extra-galactic column density $N_{H}^{ext}$ (bottom panel) 
in Figure~\ref{xrt-lc}. 
As shown in the figure, at least for 5 samples 
(GRB~050724, 060313, 070714B, 080503, and 100702A), 
the photon indices drastically change as a function of time, 
especially in the early decay phase while $N_{H}$ is rather stable. 

For example, the brightest case of GRB~050724 shows 
the spectral evolution from $\Gamma \sim 1.2$ to 3.0 
for 300~sec since the SGRB trigger time. 
GRB~070714B, 080503 and 100702A also show rapid spectral 
softening during $200-500$~sec since each SGRB trigger. 
Only one case of GRB~060313 shows gradual softening 
over $10^{4}$~sec. 

\subsection{Decay slopes of lightcurves}
\label{subsec:decay}
In Figure~\ref{xrt-lc}, the energy fluxes are estimated by 
the spectral fitting, 
although almost all previous works converted 
the photon flux to the energy flux with averaged spectral 
parameters. Then we investigated the temporal behaviour 
with three models,
i.e., single power-law (PL), broken power-law (BPL), 
and exponential (EXP) functions.
We summarizes the fitting results in Table~\ref{tbl:table2}. 
Because of small flaring activities or some fluctuations of 
the early X-ray afterglows, the reduced $\chi^2$ values are 
rather large. However the PL model is acceptable for 5 SGRBs, 
GRB~051221A, 060313, 070714B, 120804A, and 130603B.

On the other hand, we could significantly improve the fitting 
results with BPL and/or EXP models compared with PL model 
for the other 4 SGRBs, GRB~050724, 080503, 090510, and 100702A.
Especially for 3 events (GRB~050724, 080503, and 100702A),
their temporal indices after the break time are remarkably steep. 
In the remaining GRB~090510, the BPL model is better than 
the PL model, but the temporal index is gentle $1.97 \pm 0.35$ 
even after the break, which is different from 
the previous three events. 
When we adopt EXP model to the X-ray lightcurves of 
GRB~050724, 080503, and 100702A, the time constant is
obtained as $50-100$~sec (see Table~\ref{tbl:table2}). 
The redshift is measured only for GRB~050724 as $z=0.258$, 
and then the intrinsic time constant is $41.5 \pm 0.7$~sec in this case. 

\section{Discussion} \label{discussion}
\label{sec:discussion}
We systematically studied X-ray properties of SGRBs for 
the brightest 8 events observed by {\it Swift}-XRT.
We performed time resolved spectral analyses for all events,
and measured energy fluxes taking the spectral parameters in each time bin
into account. In this section, we discuss the observed X-ray 
properties of SGRBs, and classify them into two types.

\subsection{Connection between Extended X-ray Emission and Rapid Decay }
\begin{figure}
\begin{center}
\includegraphics[angle=0,scale=0.20]{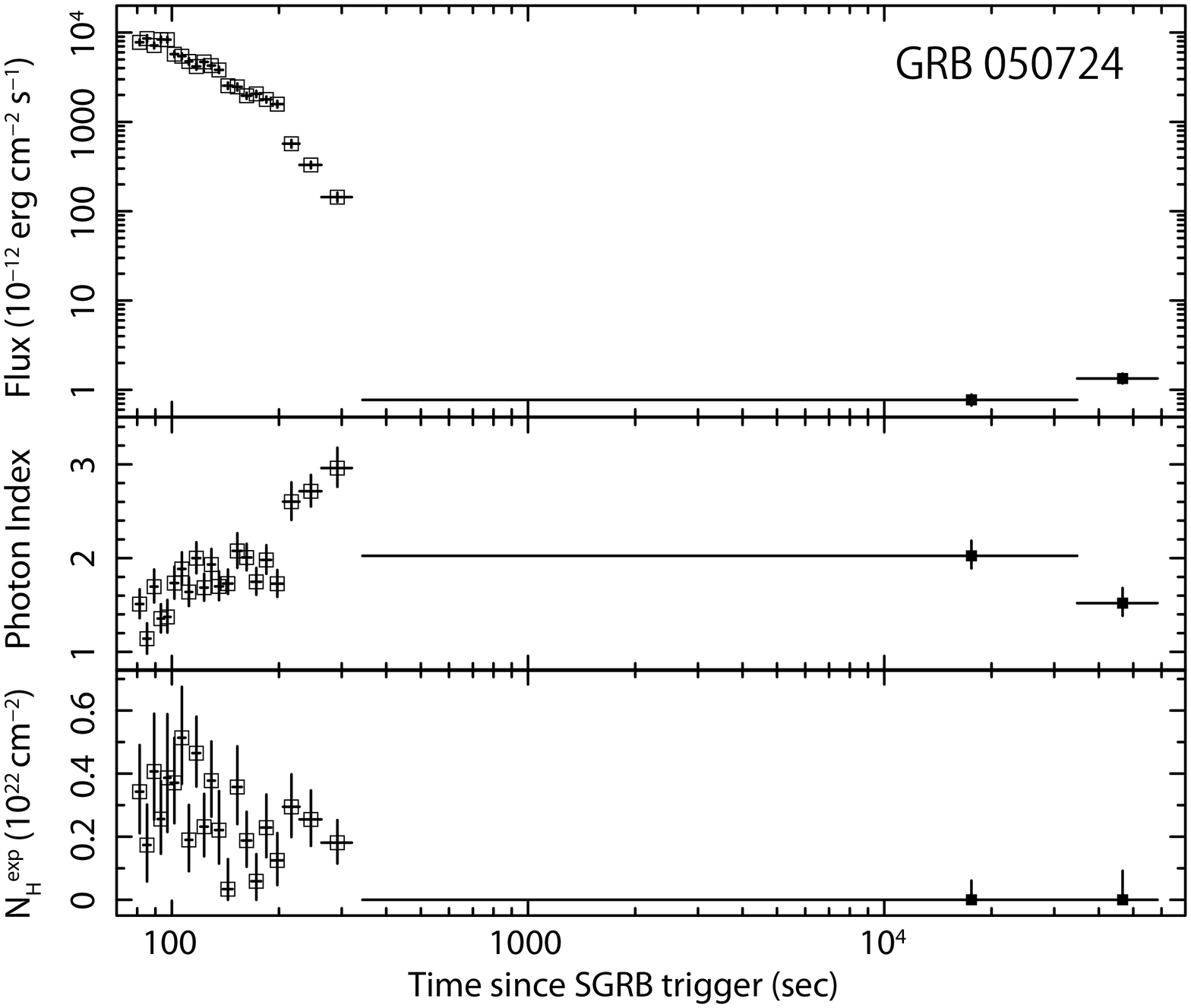}
\includegraphics[angle=0,scale=0.20]{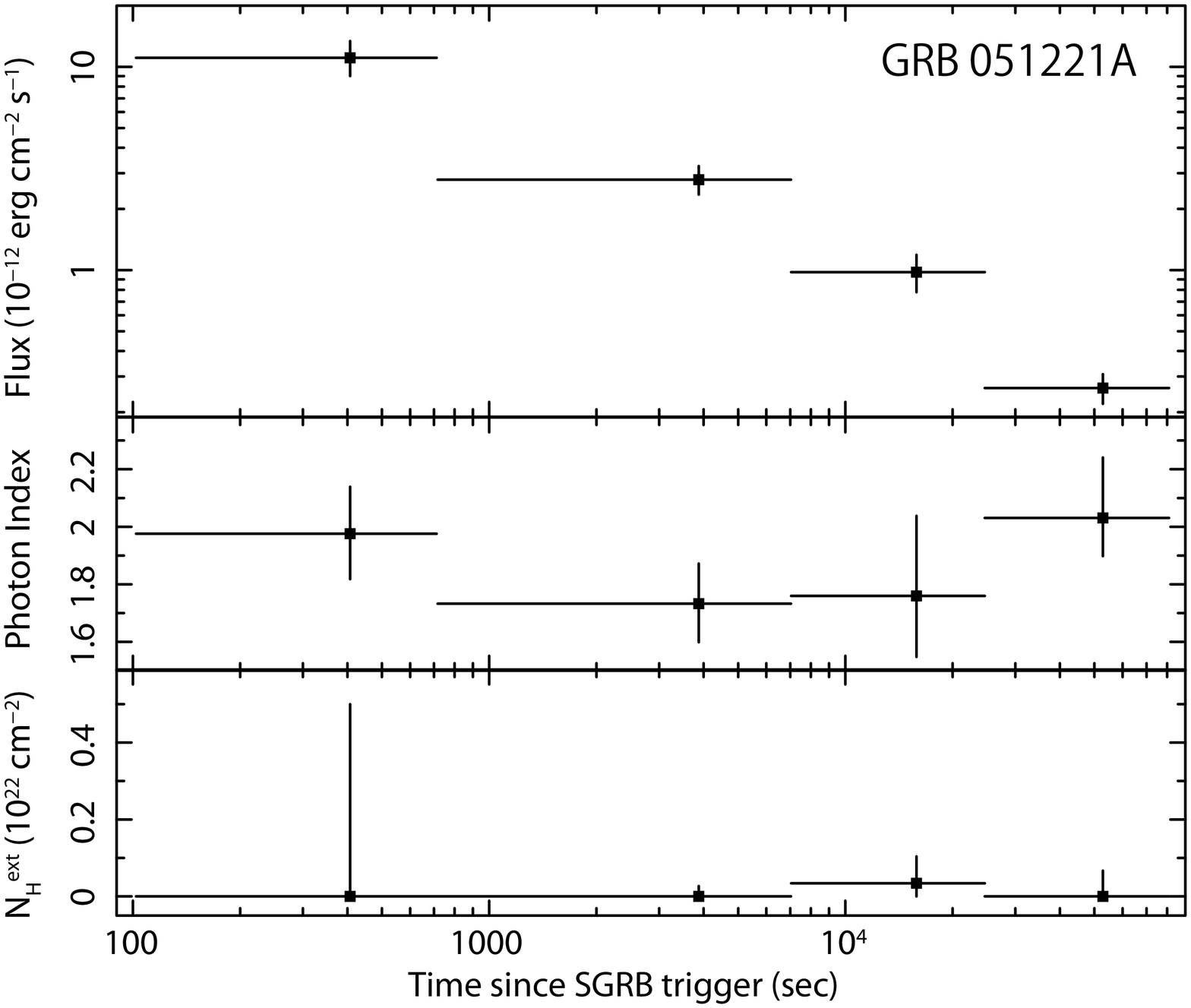}
\includegraphics[angle=0,scale=0.20]{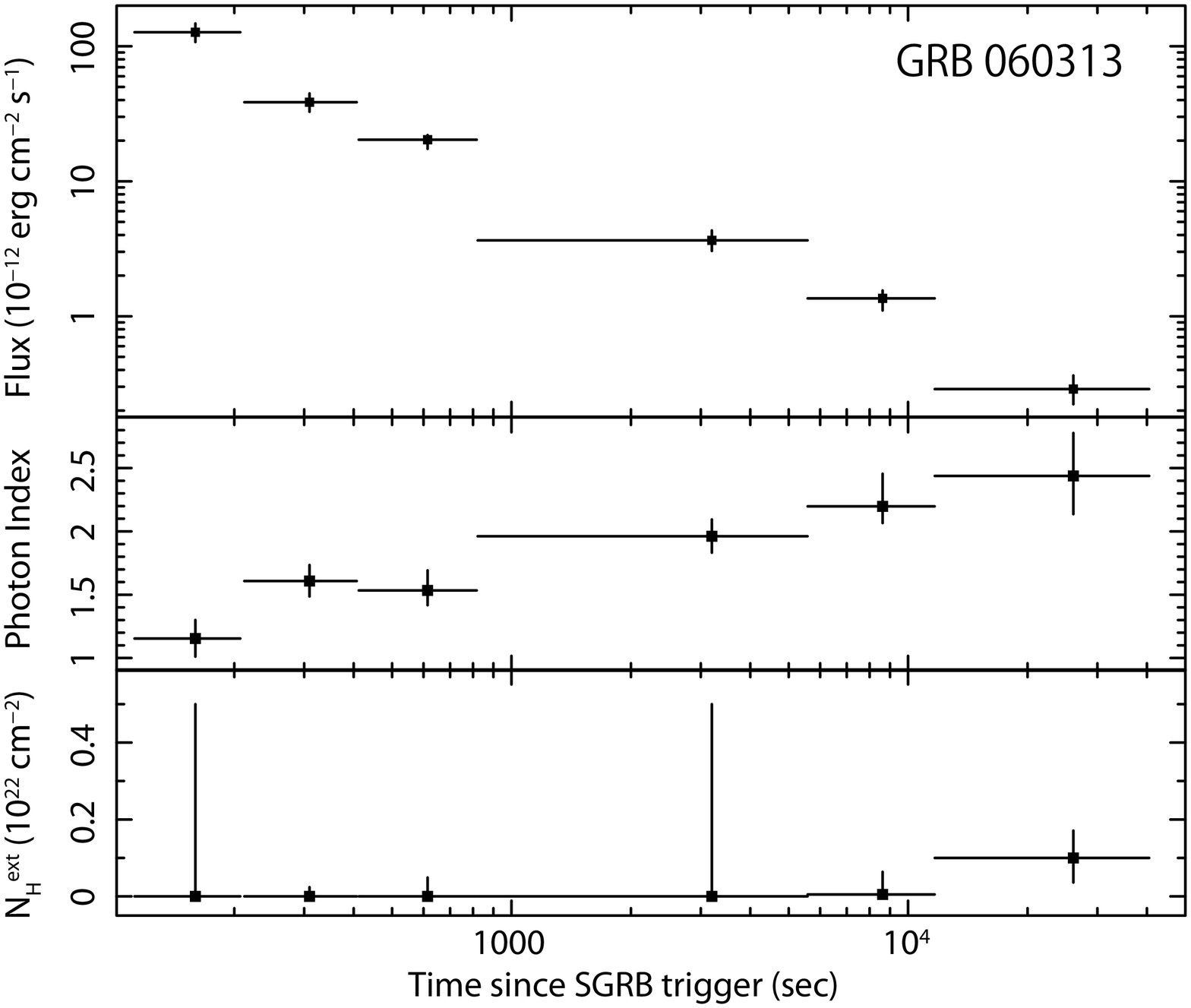}
\includegraphics[angle=0,scale=0.20]{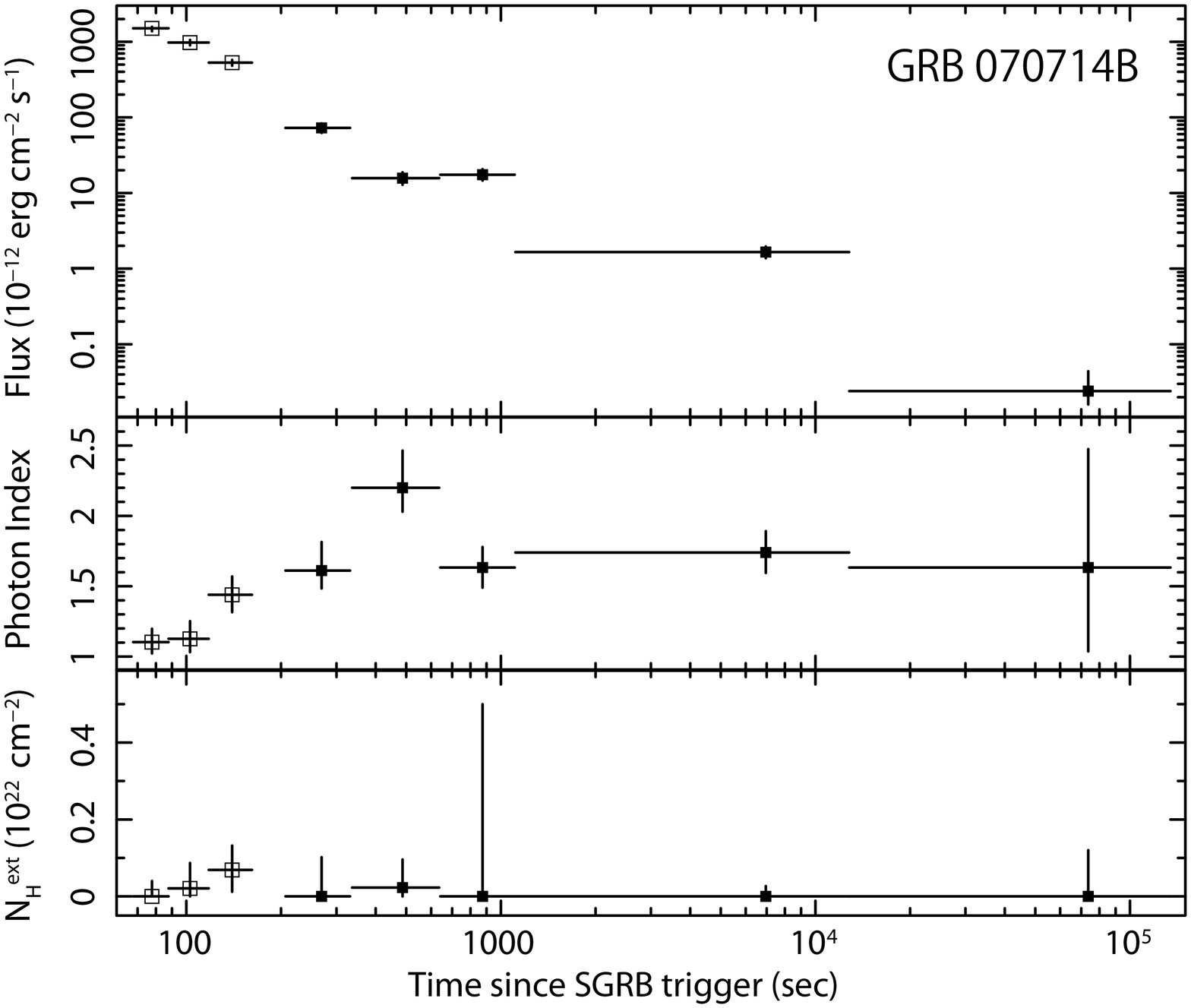}
\includegraphics[angle=0,scale=0.20]{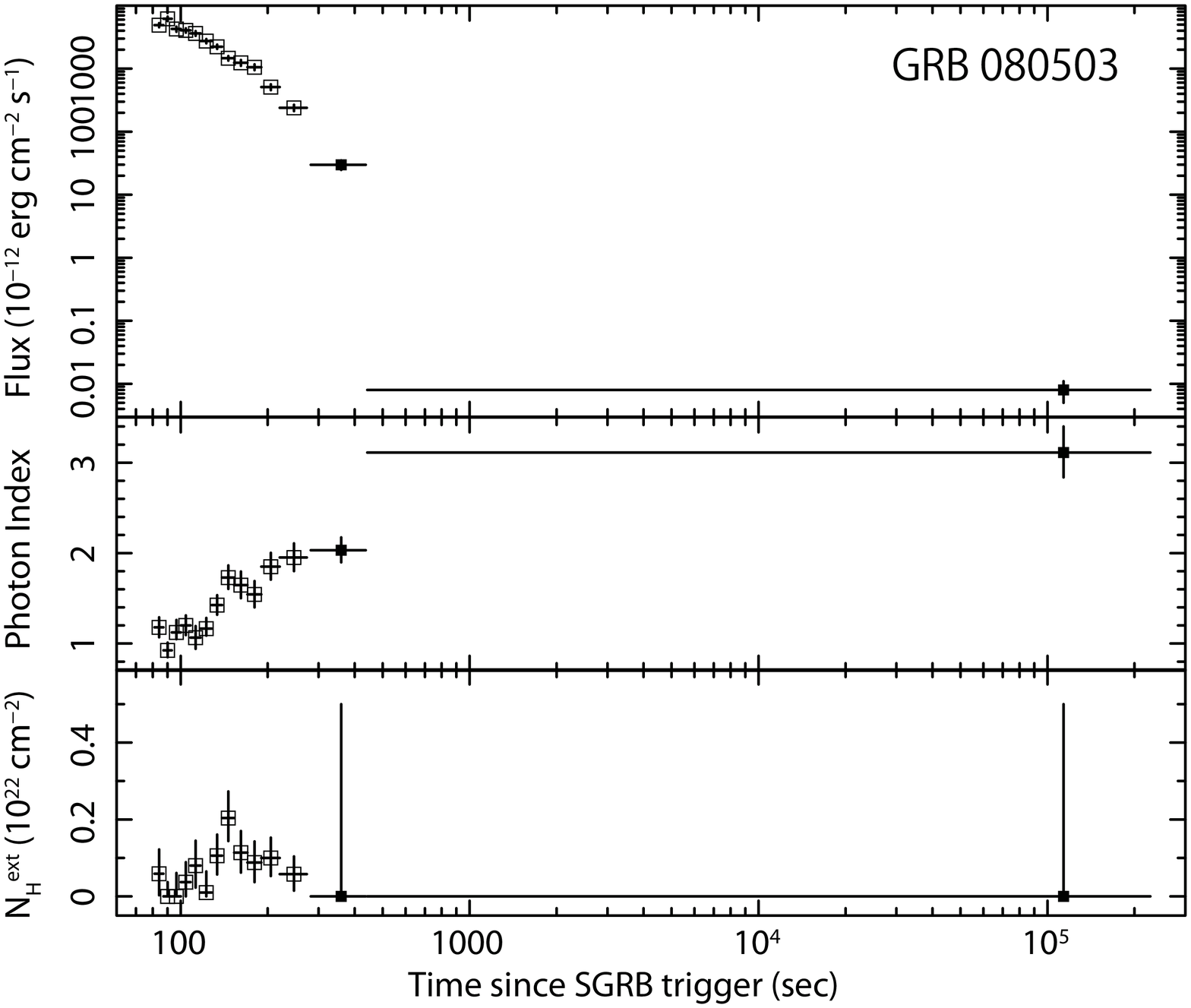}
\includegraphics[angle=0,scale=0.20]{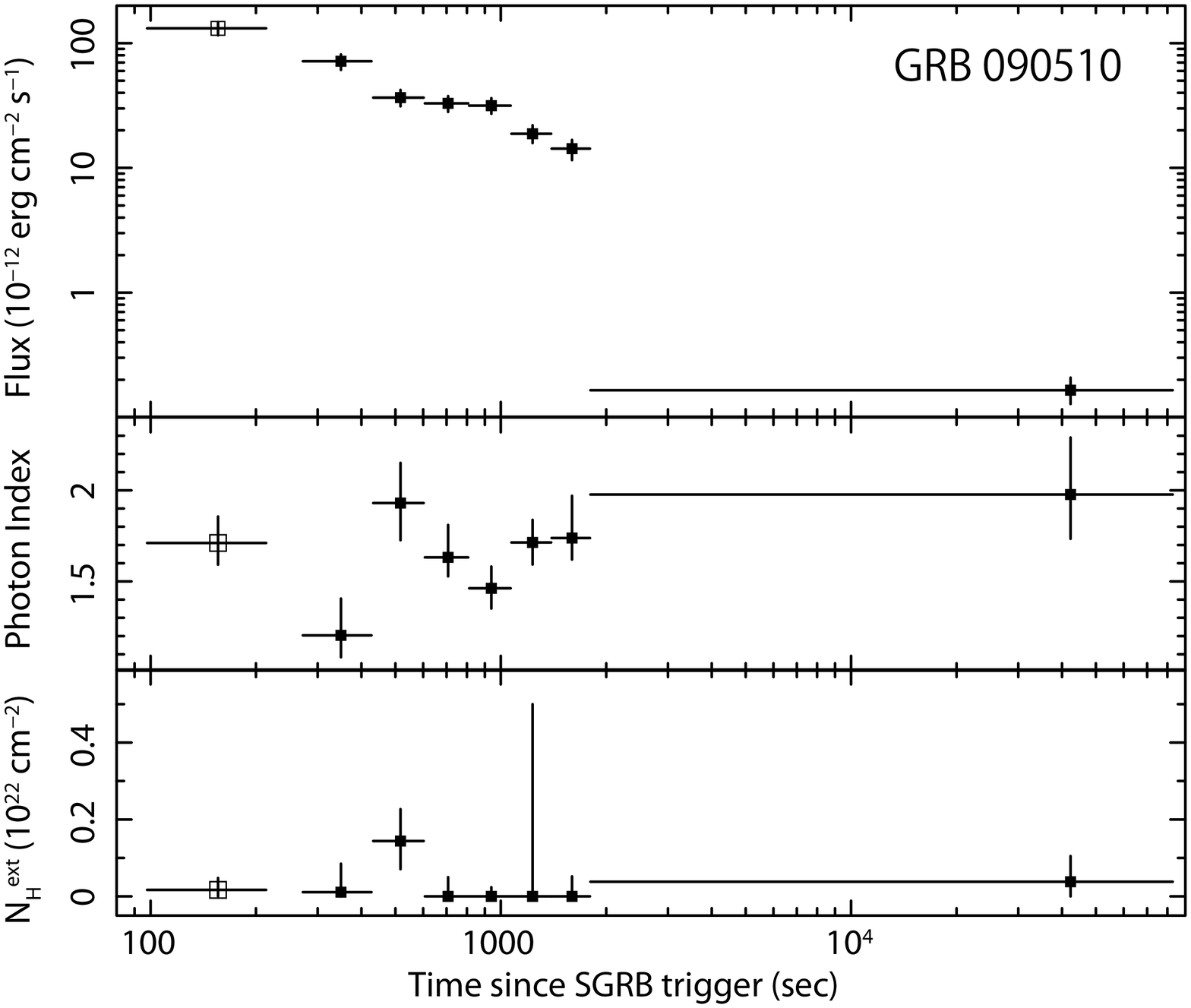}
\includegraphics[angle=0,scale=0.20]{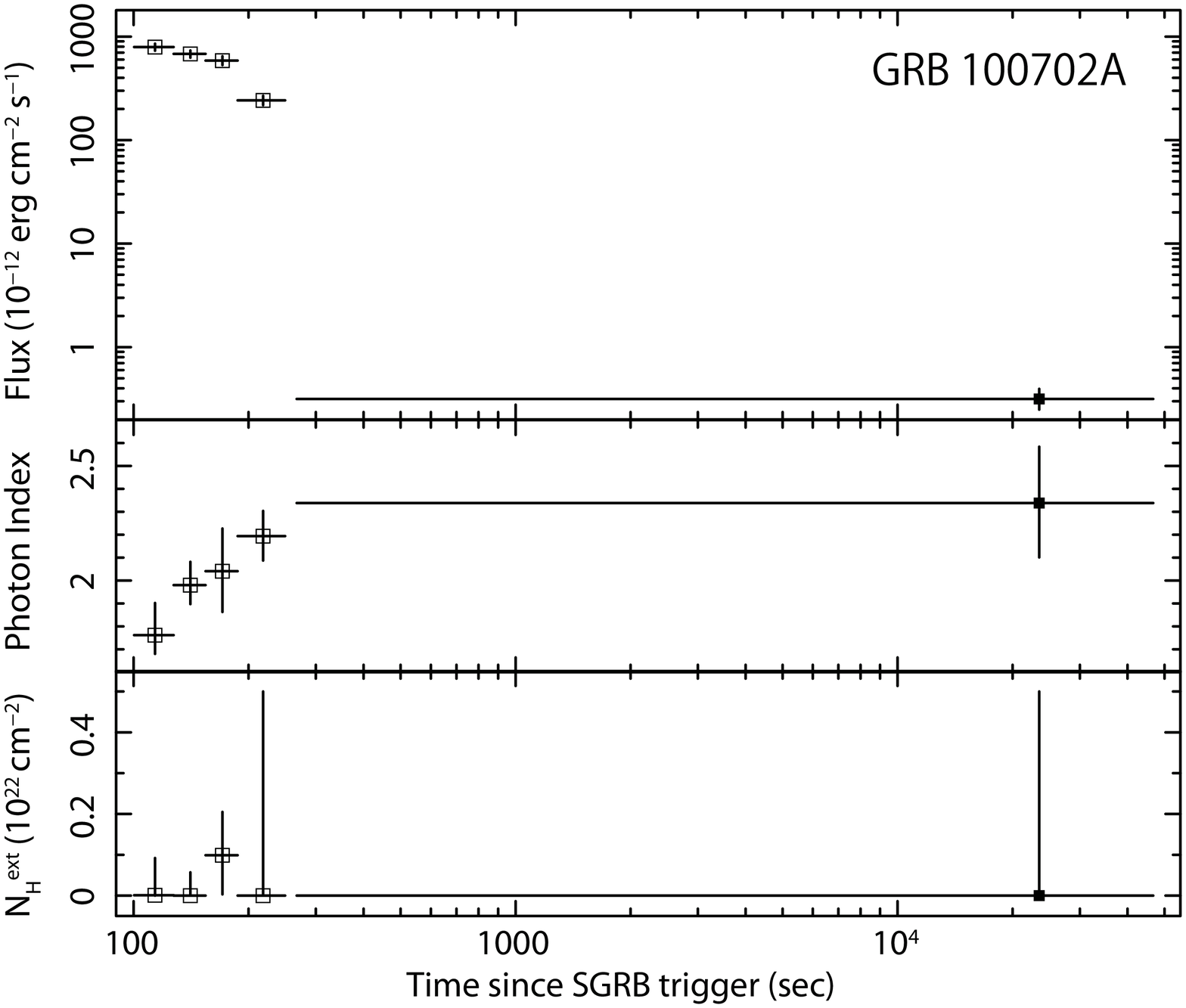}
\includegraphics[angle=0,scale=0.20]{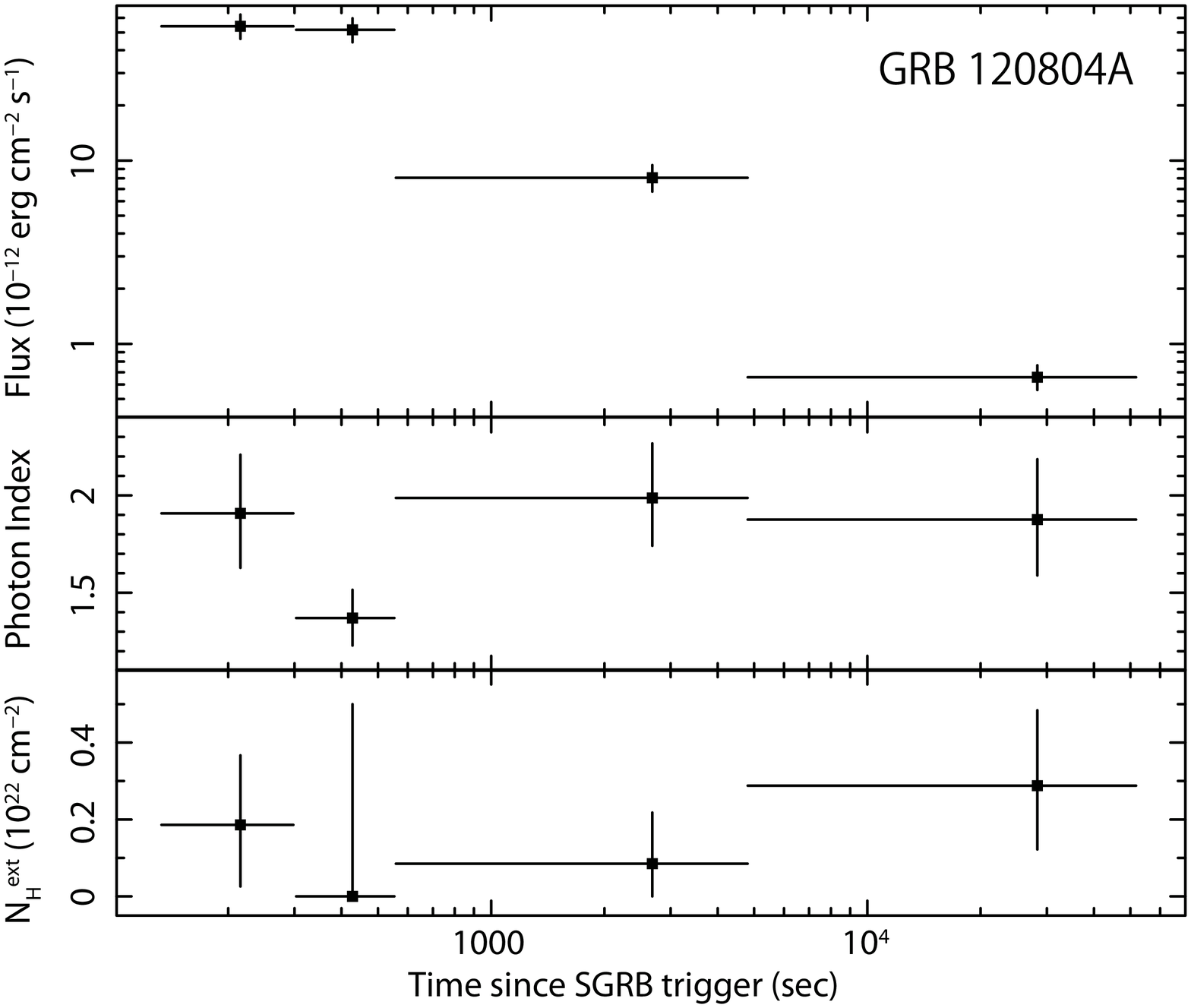}
\includegraphics[angle=0,scale=0.20]{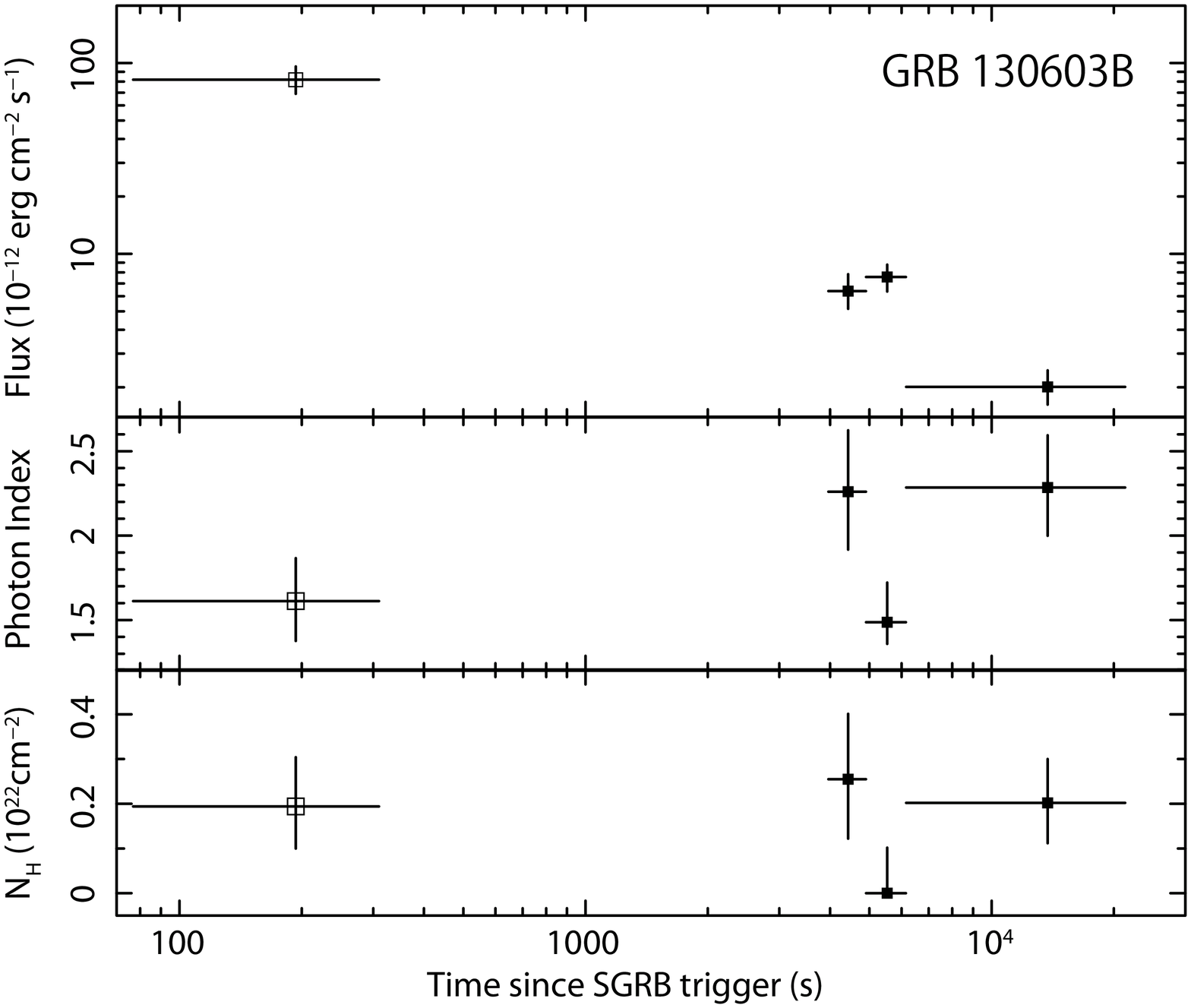}
\caption{(Top panel) Observed X-ray lightcurves by 
{\it Swift}-XRT. Open and filled squares indicate 
the data of WT mode and PC mode of {\it Swift}-XRT 
observations, 
respectively. Each energy flux in 2.0--10.0~keV is measured 
by time resolved spectral analyses. (Middle panel) The photon 
indices obtained by spectral fittings with the absorbed 
power-law model (Equation~\ref{eq:power-law}). 
(Bottom panel) The extra-galactic column density $N^{ext}_{H}$ 
in units of $10^{22}~{\rm cm^{-2}}$.
\label{xrt-lc}}
\end{center}
\end{figure}

Comparing Figure~\ref{bat-lc} and Figure~\ref{xrt-lc},
three SGRBs with the obvious extended X-ray emission 
in {\it Swift}-BAT lightcurves (GRB~050724, 070714B, and 080503) 
have the strong spetral softening and also show rapid decay for two them 
(GRB~050724, 080503).
GRB~100702A did not have 
strong extended emission in the BAT lightcurve
while it shows spectral softening. The extended emission of 
GRB~100702A is most likely under the sensitivity of {\it Swift}-BAT
since the X-ray flux in {\it Swift}-XRT is dimmer than 
the other three events in Figure~\ref{xrt-lc}. 
These results are summarized in Table~\ref{tbl:table3}.

On the other hand, the remaining 5 events 
without extended emission in {\it Swift}-BAT lightcurves
(GRB~051221A, 060313, 090510, 120804A, and 130603B) 
do not show the rapid decay phase, and their X-ray lightcurves are 
almost fully consistent with single power-law decay 
in {\it Swift}-XRT observations. Therefore we conclude that 
the extended X-ray emission in BAT lightcurves has the same origin 
as the rapid decay in XRT lightcurves,
which has the power-law spectral shape and also shows rapid spectral 
softening with time scale of $100-1000$~sec.

\subsection{Exponential Decay in Early Phase}
In \S~\ref{subsec:decay}, we adopt both BPL and EXP functions 
for the lightcurves of rapid decay phase. According to the 
reduced $\chi^{2}$ values in Table~\ref{tbl:table2}, 
it is difficult to distinguish which model is more 
appropriate function to describe the rapid decay phase. 
This is because, in general, the XRT starts the follow-up 
observations after $\sim 100$~sec since GRB triggers
and hence the lightcurves are already steeply declining.
Moreover small flaring activities disturb the baseline shape 
of early decay phase. 

In BPL model, the best fit temporal index is $-5 \sim -6$ 
after the break time. In the standard afterglow model, 
i.e. synchrotron radiation from high-energy electrons 
accelerated by a relativistic shock, 
the steepest temporal index is
$t^{(2-3p)/4}$ corresponding to the temporal evolution in 
the highest-energy spectral segment \citep{piran1999}.
Here the parameter $p$ is the power-law index of the energy 
distribution of accelerated electrons. If we assume the temporal 
index of $-5$ and $-6$, the corresponding index is 
$p = 22/3$ and $p = 26/3$, respectively. 
These are too soft to realize in the usual particle acceleration.
Therefore we should include an additional idea to describe 
the steep decay.

On the other hand, in the EXP model, the observed 
time constant seems to be 
much the same as
$50-100$~sec for all events 
with the rapid decay phase. In the case of long GRBs, 
the lightcurves of the early X-ray decay phase
can be described by the EXP model 
\citep{obrien2006, sakamoto2007, willingale2010,antonios2014, imatani2015}.
Especially, \citet{imatani2015} first reported an obvious evidence 
of the exponential decay from the prompt emission to the following 
rapid decay (prompt tail) phase of GRB~100418A in $0.7-7.0$~keV 
energy band combined with MAXI-SSC and {\it Swift}-BAT data. 
Its decay constant of $31.8 \pm 1.6~{\rm sec}$ is similar 
to the time scale of SGRBs shown in this paper. 
Therefore, 
because of the analogies
between long and short GRBs, 
the EXP model may be appropriate to describe the time 
behaviour of rapid decay in SGRBs.

\citet{yonetoku2008} and \citet{moretti2008} reported 
similar rapid decline in time with strong spectral evolution 
in long GRBs. Especially \citet{yonetoku2008} 
interpreted the temporal and spectral behavior as dynamic evolution 
of a spectral model, i.e., a spectral model of a broken power-law 
with an exponential cutoff moves through the observational energy window 
of XRT during the rapid decay phase. 
Here the physical interpretation of the break energy and 
the cutoff energy is the $E_{peak}$ 
corresponding to the minimum energy of accelerated electrons and 
the synchrotron cutoff, respectively. In this paper, we cannot 
investigate similar and detailed analyses because of limited photon 
fluxes, but the dynamic spectral evolution may be a possible 
explanation for the observed temporal and spectral 
evolution in SGRBs.

We additionally study a possibility that the start time of 
extended emission is different from the trigger time 
of short prompt emission. As shown in Figure~\ref{bat-lc},
the peak times of extended emission in BAT lightcurves 
are about 80~sec and 50~sec for GRB~050724 and GRB~080503, 
respectively. Therefore we redefine the origin of start time
to these peak times,
and measure the temporal index of rapid decay phase.
When we adopt the BPL model to the time-shifted lightcurves,
we obtained the temporal index of 
$\alpha_1 = 0.32 \pm 0.01$, $\alpha_2 = 3.72 \pm 0.14$, 
and the break time of $t_b \sim 94$~sec for GRB~050724, 
and 
$\alpha_1 = 1.22 \pm 0.07$, $\alpha_2 = 4.03 \pm 0.21$, 
and $t_b \sim 125$~sec for GRB~080503.
We can recognize these temporal indices are still steeper than 
the general GRB afterglow phenomena, and conclude that the extended
emission really shows the rapid decay in time.

\subsection{Correlation between temporal and spectral indices}
\label{subsec:alpha-beta}
According to the previous sub-sections of 
\S~\ref{subsec:xrt-spectra} and \S~\ref{subsec:softening},
both temporal and spectral indices vary as a function of time. 
When we describe the energy flux as 
$F_{\nu} \propto t^{-\alpha} \nu^{-\beta}$, 
there is a well-known correlation between $\alpha$ and $\beta$ as 
$\alpha = \beta + 2$ ($\beta = \Gamma - 1$), so called  
$\alpha$-$\beta$ correlation,
that is realized if a uniform relativistic jet suddenly stops 
the emission and the high-latitude emission dominates the flux
\citep{kumar2000}.
It is important to understand the emission mechanism and 
geometry of SGRB's emitting regions. Therefore we systematically 
investigate the $\alpha$-$\beta$ correlation. 

To estimate a gradient of the lightcurve, we first adopt 
the exponential function to describe the 
rapid decay. 
After that, we numerically calculate the exponential function 
and create the pseudo lightcurve with the best fit parameters. 
Finally, we divide the pseudo lightcurve into the same time 
intervals as in the spectral analyses, and applied the power-law spectral model 
to the time resolved pseudo lightcurve. 
By doing this, we can avoid the disturbance from fluctuations and 
small flares, and estimate the basic trend of the rapid decay phase.
(If we directly estimate the temporal index from lightcurve data, 
we frequently obtain unrealistic temporal index because of 
the rattled shape of lightcurves.)
Here, we again emphasize that the temporal 
indices are measured from lightcurves 
with time resolved 
spectral analyses in our work, while previous works 
converted 
the photon counting rate
to the energy flux with averaged spectral parameters. 

\begin{figure}
\begin{center}
\includegraphics[angle=0,scale=0.5]{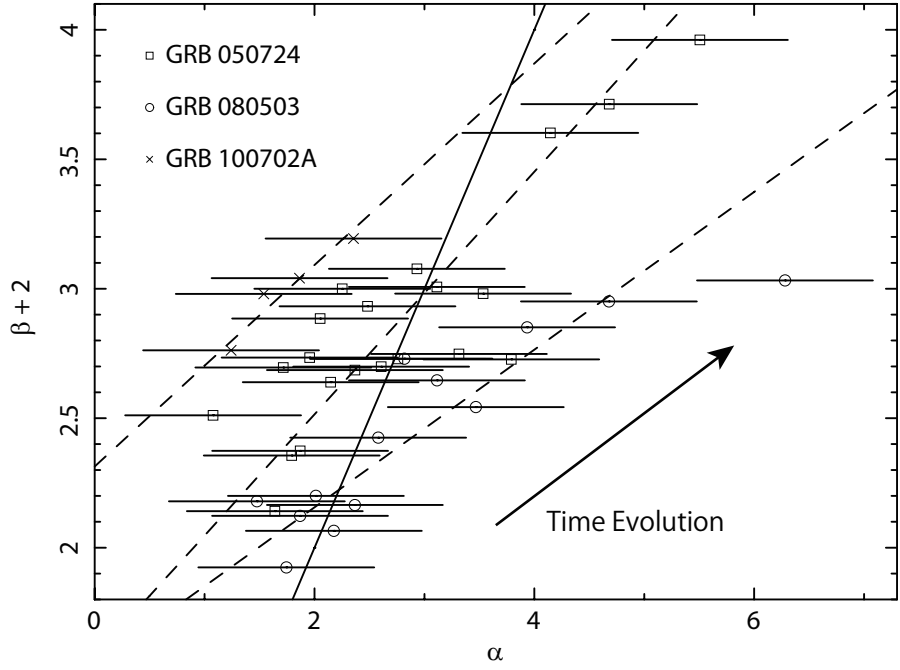}
\caption{Correlations between temporal and spectral 
indices of three SGRBs, GRB~050724 (open squares), 
080503 (open circles) and 100702A (crosses).
The solid line is the expected function of $\alpha = \beta + 2$ 
for the high-latitude emission of 
spherically symmetric shells,
and the dashed lines are the best fit linear function 
for each event. For GRB~050724 and GRB~080503, both correlations
cross the solid line at $t \sim 110~{\rm sec}$ and 
$t \sim 152~{\rm sec}$ since the SGRB trigger time, respectively.
\label{ab-plot}}
\end{center}
\end{figure}

In Figure~\ref{ab-plot}, we show $\alpha$-$\beta$ correlations 
of GRB~050724, 080503 and 100702A. The error size in the temporal 
index $\alpha$ is a representative value of 
$\Delta \alpha = \pm 0.8$ when 
we directly measure the temporal power-law index
by using neighboring three points in the observed lightcurve.
The solid line is the expected function of $\alpha = \beta + 2$ 
for high-latitude emission from a uniform jet \citep{kumar2000}. 
The dashed lines are the best fit function for each SGRB.

The best fit function is
$\beta + 2 = (0.5 \pm 0.1) \alpha + (1.6 \pm 0.3)$ for GRB~050724,
$\beta + 2 = (0.3 \pm 0.1) \alpha + (1.6 \pm 0.2)$ for GRB~080503,
and $\beta + 2 = (0.4 \pm 0.4) \alpha + (2.3 \pm 0.7)$ for GRB~100702A, 
respectively.
In these results, the slopes of the linear fits are
similar to each other, and inconsistent with the 
high-latitude emission of the slope 1 with more than 3~$\sigma$ 
statistical level. The intercept values of the linear fits are 
different from each other.

Finally, let us argue possible implications of the observed 
$\alpha$-$\beta$ correlation for the SGRB models.
The extended emission is most likely caused by the long-lasting 
activity of the central engine because the rapid decay is very steep 
while the external shock emission can not generally produce such large 
variabilities in the afterglow lightcurves \citep{ioka+05}
\footnote{If the density bump of the circumburst medium 
decelerates the jet down to the Lorentz factor less than 
the inverse of the opening angle $\Gamma < \theta_j^{-1}$, 
the light curve might decline steeply, although 1 dimensional 
simulations are not conclusive\citep{mimica2011,mesler2012}.}.
Then the rapid decay phase at $\delta t \sim 100 \ \rm sec$ 
signals the quenching of the jet from the central engine.

The important point of our finding is that the temporal decay 
of the extended emission becomes even faster than those produced 
by the high-latitude emission from quenching uniform jets,
as one can see from  Figure~\ref{ab-plot}.
This means that the contribution from the high-latitude emission 
is smaller than that of a uniform jet.
Therefore the emission geometry of the jet should not be uniform; 
the brightness declines significantly outside the solid angle of
$\theta_{jet} \gtrsim 1/\Gamma \sim 0.01$ on the line-of-sight,
where $\Gamma$ is the bulk Lorentz factor of the jet
and is larger than $\sim 100$ to avoid the compactness problem.
Although this may imply the jet opening angle is small or patchy
in the extended emission and rapid decay phase \citep{yamazaki+04},
a typical opening angle of a SGRB jet is usually
$\theta \sim 0.1$, much larger than $1/\Gamma$ 
\citep{fong+14,nagakura+14,nakamura2014,mizuta+13}.
A natural solution to this inconsistency may be that 
the emission comes from a photosphere \citep{rees+05,ioka+07,peer+12}.
The photospheric radius becomes large outside 
the viewing angle of $1/\Gamma$ for a relativistic jet 
because the photosphere is concave for $v/c > 2/3$ while it is
convex for $v/c < 2/3$ as shown by \citep{abramowicz+91}.
Therefore the high-latitude emission from the photosphere is 
suppressed even if the jet is uniform.

Additionally, we investigate the $\alpha$-$\beta$ correlation
for the time shifted lightcurve of GRB~050724 and GRB~080503.
We exclude GRB~100702 because it does not show the obvious 
extended emission in BAT lightcurve and we can not determine 
its start time as shown in Figure~\ref{bat-lc}.
Then we estimated the temporal index of $\alpha$ with the
same method as explained above except for the time shift
of 80~s and 50~s for GRB~050724 and 080503, respectively.

\begin{figure}
\begin{center}
\includegraphics[angle=0,scale=0.5]{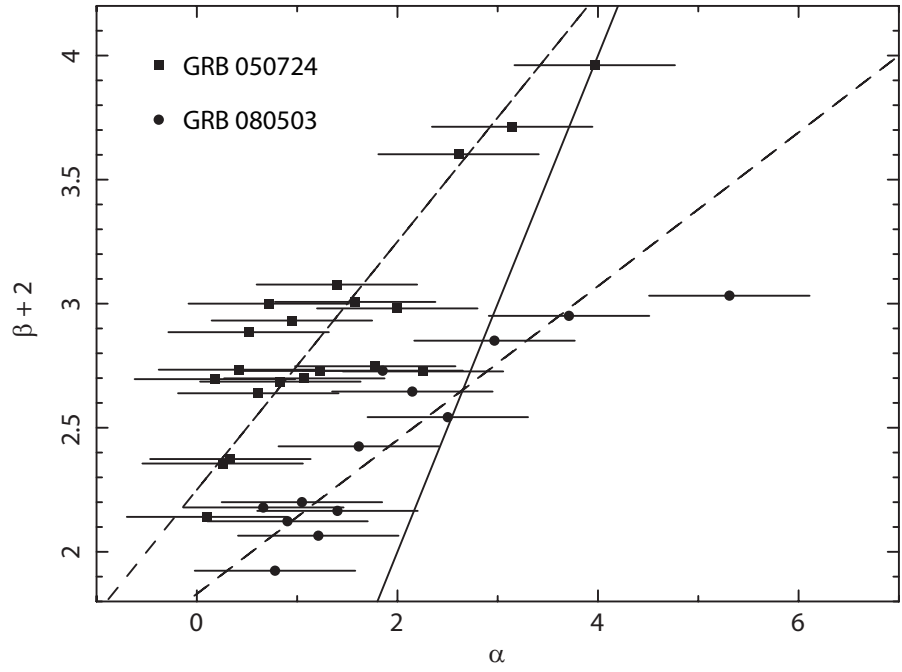}
\caption{Correlations between temporal and spectral 
indices of GRB~050724 (filled squares) and 
080503 (filled circles) after shifting the time origin by 80~s and 50~s, respectively.
The meanings of solid and dashed lines are the same as
Figure~\ref{ab-plot}.
\label{ab-shifted}}
\end{center}
\end{figure}

In Figure~\ref{ab-shifted}, we again show the $\alpha$-$\beta$
correlation for the time shifted data. The best fit function is
$\beta + 2 = (0.5 \pm 0.1) \alpha + (2.3 \pm 0.2)$ for GRB~050724,
and $\beta + 2 = (0.3 \pm 0.1) \alpha + (1.8 \pm 0.1)$ for GRB~080503. 
The slopes of these results are consistent with 
the previous ones before the time shift, and still inconsistent 
with the slope of 1 expected from the high-latitude emission.
Even if we consider the time shifted lightcurves, 
at least one data point is still clearly in the region of $\alpha > \beta +2$
as shown in Figure~\ref{ab-shifted}.
Therefore the above discussions may be also
adopted in this case.

The observed $\alpha$-$\beta$ correlation may reflect
the angular structure of the photosphere. It is an interesting 
problem to study whether the $\alpha$-$\beta$ correlation is
reproduced in the photosphere model or not.
Alternatively, the observed $\alpha$-$\beta$ correlation may be 
directly produced by the declining jet emission. 
In the black hole model with fallback accretion, the Blandford-Znajek 
process is the most likely mechanism to launch a relativistic jet
\citep{BZ77},
and the rapid decay in X-ray corresponds to the magnetic field 
decay via reconnection at the black hole, which reduces the energy 
extraction from the rotating black hole \citep{kisaka2015}.
Although currently we cannot predict the spectral evolution,
our observations should give a clue to the unknown mechanism of 
the jet emission.

\subsection{Population statistics of extended emission}

In this paper, we selected the brightest 9 SGRBs in early X-ray flux
observed by {\it Swift}-XRT. The 8 of them are also the brightest
events in fluence of BAT observation including the extended emission
if it exists, except for 3 SGRBs with the observation start time
of $\sim 200$~s which is later than the usual.\footnote{GRB~101219A, 140930B, and 080426 are excluded.}
These 8 events are satisfied with the criteria of
BAT fluence of $>3.4 \times 10^{-7}~{\rm erg~cm^{-2}}$ and
XRT flux of $> 10^{-11}~{\rm erg~cm^{-2}s^{-1}}$ from selected
92 SGRB candidates. In our sample, only GRB~100702A is the outlier of
dimmer fluence in BAT observation but brighter in XRT.

Focusing on these 8 events, the ratio of SGRBs with extended emission
is 37.5~\%. \citet{Bostanci2013} found 7~\% of SGRBs have
the extended emission in BATSE data, and \citet{Norris2010}
and \citet{sakamoto2011} reported 25~\% and 2~\% in BAT data,
respectively. These statistical values should not be directly compared
because our and their criteria are different from each other.
But, in this paper, we found that the early X-ray flux observed by
XRT is strongly affected by the long lasting tail of extended emission.
Therefore, our criteria including the early X-ray flux will be
important for the future investigation.

\subsection{Conclusions}
In conclusion, we find the following properties of SGRBs:
\begin{enumerate}
\item The spectra of the extended soft X-ray emission and 
following rapid decay phase can be described by a power-law function 
with spectral softening. We can 
exclude a simple 
thermal blackbody function.
\item The rapid decay phase 
usually follows the extended X-ray emission.
\item The X-ray lightcurve from the extended emission to 
the rapid decay phase can be fitted by the exponential function 
with the time constant of $50-100$~sec.
\item The high latitude emission can not explain the temporal and 
spectral behaviour of the extended emission and rapid decay, 
because 
the observations do not follow the
expected $\alpha$-$\beta$ correlation ($\alpha = \beta + 2$).
\item The extended X-ray emission may be observed in 37.5~\% of bright SGRBs
on the selection criteria of BAT fluence of
$>3.4 \times 10^{-7}~{\rm erg~cm^{-2}}$ and
XRT flux of $> 10^{-11}~{\rm erg~cm^{-2}s^{-1}}$.

\end{enumerate}


\section*{Acknowledgments}
This work is supported in part by the Grant-in-Aid from the Ministry of 
Education, Culture, Sports, Science and Technology (MEXT) of Japan, 
No. 25103507, 25247038, 15H00780 (DY), 
No. 24103006, 15H02087 (TN), 
No. 24340048, 26610048 (KT),
No. 26287051, 24103006, 24000004, 26247042 (KI).
KK is supported by NASA through Einstein Postdoctoral Fellowship 
grant number PF4-150123 awarded by the Chandra X-ray Center, 
which is operated by the Smithsonian Astrophysical Observatory 
for NASA under contract NAS8-03060.

\clearpage
\begin{deluxetable}{llrrc} 
\tablewidth{0pc} 
\tablecaption{Samples of Short GRB \label{tbl:table1}} 
\tablehead{ 
\colhead{ID} & \colhead{redshift} & \colhead{$T_{90, obs}$ (sec)} &
 \colhead{$\chi^2$(d.o.f) of E.E.}& \colhead{reference}}
\startdata 
GRB~050724  & 0.258$^a$  & 98.7  &  105.3 (23)& $^a$\citet{prochaska2005}\\
GRB~051221A & 0.5465$^b$ &  1.4  &   17.9 (23)& $^b$\citet{berger2005}\\
GRB~060313  & ---    &  0.8  &   21.9 (23)& \\
GRB~070714B & 0.92$^c$   & 65.2  &   59.9 (23)& $^c$\citet{graham2007}\\
GRB~080503  & ---    & 274.9 &  712.8 (23)& \\
GRB~090510  & 0.903$^d$  &  0.4  &   23.2 (23)& $^d$\citet{rau2009}\\
GRB~100702A & ---    &  0.2  &   18.3 (23)& \\
GRB~120804A & ---    &  1.8  &   23.2 (23)& \\
GRB~130603B & 0.3586$^e$ &  0.18  &   14.5 (23)& $^e$\citet{Cucchiara2013}
\enddata 
\end{deluxetable} 

\begin{deluxetable}{lccccccc}
\tablewidth{0pc}
\tablecaption{Temporal properties of early X-ray emission 
(extended emission or afterglow) of SGRBs
\label{tbl:table2}}
\tablehead{
\colhead{ID} & \colhead{model} & \colhead{index} &
 \colhead{break time} & \colhead{index} &
 \colhead{time const.} &  \colhead{$\chi^2$} & \colhead{d.o.f} \\
\colhead{} & \colhead{} & \colhead{$\alpha_1$} &
 \colhead{$t_b$} & \colhead{$\alpha_2$} &
 \colhead{$\tau$} &  &
}
\startdata
GRB~050724  & PL  & $2.63 \pm 0.19$ & --- & --- & --- & 280 & 19\\
            & BPL & $2.03 \pm 0.07$ & 187 & $5.90 \pm 0.53$ & --- & 55.2 &
17\\
            & EXP & --- & --- & --- &  $52.2 \pm 0.9$ &  81.7 & 19\\
\hline
GRB~051221A & PL  & $0.77 \pm 0.04$ & --- & --- & --- & 4.0 & 2\\
            & BPL & $0.66 \pm 0.03$ & 25700 & $1.47 \pm 0.24$ & --- & 0.3 &
0 \\
            & EXP & --- & --- & --- &$ 192 \pm 39 $ & 103 & 2\\
\hline
GRB~060313  & PL  & $1.11 \pm 0.04$ & --- & --- & --- & 5.5 & 4\\
            & BPL & $1.07 \pm 0.03$ & 20100 & $2.07 \pm 0.37$ & --- & 4.9 &
2\\
            & EXP & --- & --- & --- & $305 \pm 60$ & 93.7 & 4\\
\hline
GRB~070714B & PL  & $2.33 \pm 0.06$ & --- & --- & --- & 76.2 & 6\\
            & BPL & $1.58 \pm 0.17$ & 117 & $2.64 \pm 0.14$ & --- & 67.7 &
4\\
            & EXP & --- & --- & --- & $68.3 \pm 4.0$ & 100 & 6\\
\hline
GRB~080503  & PL  & $3.34 \pm 0.22$ & --- & --- & --- & 143 & 11\\
            & BPL & $2.27 \pm 0.12$ & 179 & $5.06 \pm 0.29$ & --- & 24.4 &
9\\
            & EXP & --- & --- & --- &  $51.9 \pm 1.3$ &  18.3 & 11\\
\hline
GRB~090510  & PL  & $1.17 \pm 0.03$ & --- & --- & --- & 19.0 & 6\\
            & BPL & $0.93 \pm 0.01$ & 10200 & $1.97 \pm 0.35$ & --- & 4.7 &
4\\
            & EXP & --- & --- & --- & $706 \pm 87$ & 39.1 & 6\\
\hline
GRB~100702A & PL  & $1.72 \pm 0.16$ & --- & --- & --- & 13.2 & 2\\
            & BPL & $0.74 \pm 0.05$ & 192 & $6.13 \pm 0.30$ & --- & 0.002 &
0\\
            & EXP & --- & --- & --- & $91.1 \pm 8.6$ & 8.0 & 2\\
\hline
GRB~120804A & PL  & $0.93 \pm 0.04$ & --- & --- & --- & 8.9 & 2\\
            & BPL & $0.78 \pm 0.03$ & 14800 & $1.91 \pm 0.24$ & --- & 5.3 &
0\\
            & EXP & --- & --- & --- & $1260 \pm 140$ & 46.2 & 2\\
\hline
GRB~130603B & PL  & $0.83 \pm 0.05$ & --- & --- & --- & 4.72 & 2\\
            & BPL & $0.75 \pm 0.02$ & 8880 & $1.89 \pm 0.33$ & --- & 1.57 &
0\\
            & EXP & --- & --- & --- & $2130 \pm 300$ & 31.7 & 2
\enddata
\end{deluxetable}

\clearpage
\begin{deluxetable}{lccc} 
\tablewidth{0pc} 
\tablecaption{Status of Early X-ray Property of SGRBs \label{tbl:table3}} 
\tablehead{ 
\colhead{ID} & \colhead{extended} & \colhead{rapid} & \colhead{spectral} \\
    & \colhead{emission} & \colhead{decay} & \colhead{evolution}
}
\startdata 
GRB~050724  & YES       & YES & YES  \\
GRB~051221A & NO        & NO  & NO   \\
GRB~060313  & NO        & NO  & YES  \\
GRB~070714B & YES       & NO  & YES  \\
GRB~080503  & YES       & YES & YES  \\
GRB~090510  & NO        & NO  & NO   \\
GRB~100702A & NO$^{a}$  & YES  & YES  \\
GRB~120804A & NO        & NO  & NO  \\ 
GRB~130603B & NO        & NO  & NO  \\
\enddata \\
{$^{a}$ The flux level of extended emission may be under the sensitivity of BAT}
\end{deluxetable}

\clearpage

\clearpage

\clearpage

\clearpage

\clearpage

\clearpage

\end{document}